\documentclass[aps,onecolumn,preprintnumbers,showpacs,showkeys,nofootinbib,prd,
superscriptaddress  
]{revtex4}

\usepackage[american]{babel}
\usepackage[T1]{fontenc} 
\usepackage[utf8]{inputenc}
\usepackage{tikz}
\usepackage{pgfplots}
\usepackage{pgfplotstable}
\usepackage{pgffor}

\usepackage{graphicx}
\usepackage{amssymb,amsmath,amsfonts,amsthm,graphicx,psfrag}
\usepackage{hyperref}
\hypersetup{
    colorlinks=true,
    linkcolor=red,
    filecolor=magenta,      
    urlcolor=blue,
}

\usepackage{subfigure}
\usepackage{csquotes}

\newcommand{\beq}{\begin{eqnarray}}
\newcommand{\eeq}{\end{eqnarray}}

\newcommand{\Tr}{\mbox{ Tr}}

\newcommand{\mc}[1]{\multicolumn{1}{c}{#1}}

\usepackage[normalem]{ulem}

\begin{document}
\preprint{}

\title{Influence of relativistic rotation on the confinement/deconfinement transition in gluodynamics}
\author{V.\,V.~Braguta}
\email{vvbraguta@theor.jinr.ru}
\affiliation{ Bogoliubov Laboratory of Theoretical Physics, Joint Institute for Nuclear Research, Dubna, 141980 Russia }
\affiliation{ National University of Science and Technology MISIS, Leninsky Prospect 4, Moscow, 119049 Russia }
\affiliation{ Moscow Institute of Physics and Technology, Dolgoprudny, 141700 Russia }
\author{A.\,Yu.~Kotov}
\email{a.kotov@fz-juelich.de} 
\affiliation{ J\"ulich Supercomputing Centre, Forschungszentrum J\"ulich, D-52428 J\"ulich, Germany }
\author{D.\,D.~Kuznedelev}
\email{scope.denis@mail.ru}
\affiliation{ Moscow Institute of Physics and Technology, Dolgoprudny, 141700 Russia }
\author{A.\,A.~Roenko}
\email{roenko@theor.jinr.ru} 
\affiliation{ Bogoliubov Laboratory of Theoretical Physics, Joint Institute for Nuclear Research, Dubna, 141980 Russia }

\date{\today}

\begin{abstract}
In this paper we consider the influence of relativistic rotation on the confinement/deconfinement transition in gluodynamics within lattice simulation.
We perform the simulation in the reference frame which rotates with the system under investigation, where rotation is reduced to external gravitational field. 
 To study the confinement/deconfinement transition the Polyakov loop and its susceptibility are calculated for various lattice parameters and the values of angular velocities which are characteristic for heavy-ion collision experiments. Different types of boundary conditions (open, periodic,  Dirichlet) are imposed in directions, orthogonal to rotation axis. Our data for the critical temperature are well described by a simple quadratic function $T_c(\Omega)/T_c(0) = 1 + C_2 \Omega^2$ with $C_2>0$ for all boundary conditions and all lattice parameters used in the simulations. From this we conclude that the critical temperature of the confinement/deconfinement transition in gluodynamics increases with increasing angular velocity. This conclusion does not depend on the boundary conditions used in our study and we believe that this is universal property of gluodynamics. 
\end{abstract}


\maketitle

\section{Introduction}\label{sec:intro}

Recently the study of various physical systems under rotation has become 
relevant and extremely interesting research area. Rotating physical systems frequently appear in astrophysics~\cite{Watts:2016uzu,Grenier:2015pya}. Relativistic fermions with angular momentum can be 
realized in condensed matter physics~\cite{Basar:2013iaa, Landsteiner:2013sja}. It is believed that rapidly rotating quark-gluon plasma is created in heavy-ion collision experiments~\cite{Jiang:2016woz, Becattini:2007sr, Baznat:2013zx, STAR:2017ckg}. In the last example non-central heavy ion collisions generate nonzero angular momentum. Partly this angular momentum is taken away by spectator partons, but considerable part is transferred to quark-gluon plasma, created in the collision. The experimental results for $\Lambda$, $\bar \Lambda$ baryons polarization confirm this expectation and give the following average value for the angular velocity  $\Omega \sim 6$~MeV~\cite{STAR:2017ckg}. Hydrodynamic simulations of heavy-ion collisions predict even larger magnitudes of the angular velocity $\Omega \sim (20-40)\,\mbox{MeV}$~\cite{Jiang:2016woz}. These values of angular velocity lead to relativistic rotation of quark-gluon plasma. 

Rotation gives rise to lots of interesting phenomena which can be observed in heavy-ion collision experiments. For instance, chiral vortical effect~\cite{Vilenkin:1979ui, Kharzeev:2015znc, Prokhorov:2018qhq, Prokhorov:2018bql} and polarization of different particles~\cite{Rogachevsky:2010ys, Teryaev:2017wlm} are examples of such phenomena. 
In addition, relativistic rotation is believed to influence phase transitions in QCD what also can be observed in the experiments.  

There are a lot of theoretical papers dedicated to the phase transitions in rotating QCD matter (see, for instance,~\cite{Ebihara:2016fwa, Chernodub:2016kxh, Jiang:2016wvv, Zhang:2018ome, Wang:2018sur, Chernodub:2020qah, Chen:2020ath, Fujimoto:2021xix}). Mostly these studies are carried out within Nambu–Jona-Lasinio model~(NJL)~\cite{Nambu:1961tp, Nambu:1961fr} and they are focused on influence of rotation on the chiral symmetry breaking/restoration transition in QCD. Despite variety of details and results of there papers, there is one common result: rotation suppresses the chiral condensate, which leads to the decrease of the critical temperature with rotation. An interesting physical explanation of this suppression was proposed in paper~\cite{Jiang:2016wvv}. The idea is that  rotation induces a polarization effect which aligns all microscopic angular momenta along the total angular momentum. So the states with nonzero angular momentum/spin are more preferred than the states with zero angular momentum/spin which results into the suppression of the chiral condensate exposed to the rotation. 

An important disadvantage of all works based on the NJL model is that they take into account only the quark degrees of freedom whereas the gluon sector of the theory, which is responsible for the confinement, is integrated out. For this reason it is difficult to study confining properties as well as confinement/deconfinement transition within such approaches. For QCD matter under rotation this disadvantage might be crucial since gluons have spin-1 and one can expect that relativistic rotation induces polarization of gluons. The properties of QCD matter with polarized gluons might be different as compared to that without polarization. In particular, rotation can affect the confinement/deconfinement transition. So, in order to understand the impact of rotation to the phase transitions in QCD matter, one needs to apply an approach which takes into account both quark and gluon degrees of freedom. The papers~\cite{Chen:2020ath, Chernodub:2020qah, Fujimoto:2021xix} are focused on the confinement/deconfinement transition in rotating QCD. The authors of~\cite{Chen:2020ath} applied holographic approach. The author of~\cite{Chernodub:2020qah} studied rotating compact QED which also possesses confinement/deconfinement transition. The hadron resonance gas model was adopted in paper~\cite{Fujimoto:2021xix}. The results of these works indicate that the critical temperature decreases with the angular velocity. Although the results of different phenomenological studies  give interesting insights into QCD properties, we believe that lattice simulation of QCD is the most appropriate to study how rotation influences the confinement/deconfinement transition.

This paper is devoted to the study of SU(3) gluodynamics properties under rotation within lattice simulation. In particular, we are going to address the question how  rotation affects the confinement/deconfinement transition in gluodynamics. It is worth mentioning that the first lattice study of rotating QCD matter was carried out in~\cite{Yamamoto:2013zwa}, but the impact of rotation on QCD phase transition was not considered in this paper. In our paper we will use lattice formulation of rotating gluodynamics developed in~\cite{Yamamoto:2013zwa}. We would like also to note that our first results devoted to thermodynamic properties of rotation gluodynamics were published in~\cite{Braguta:2020biu}. In this paper we continue our study. 

This paper is organized as follows. Next section is devoted to the theoretical background of our study. In particular, we discuss Yang-Mills theory in external gravitational field, write its discretized action and describe boundary conditions used in our study. In Section~\ref{sec:Tc} the results of our study obtained with different boundary conditions are presented. In last section we discuss the results and draw a conclusion. In addition in Appendix~\ref{sec:app-Tc0} the influence of finite volume effects on thermodynamic properties of gluodynamics is studied. In Appendix~\ref{sec:app-list} we show lattice parameters used in our simulations. 

\section{Theoretical background}\label{sec:theory}

\subsection{Thermodynamic ensemble in presence of rotation}

To study the influence of rotation on gluodynamics properties we are going to use the approach proposed in papers \cite{Ebihara:2016fwa, Chernodub:2016kxh, Jiang:2016wvv, Zhang:2018ome, Wang:2018sur, Yamamoto:2013zwa}. The idea is to carry out the study in the reference frame which rotates with the system. Below it will be assumed that the system rotates around $z$ axis. In this reference frame there appears an external gravitational field with well known metric tensor
\begin{equation}\label{eq:metric}
g_{\mu \nu} = 
\begin{pmatrix}
1 - r^2 \Omega^2 & \Omega y & -\Omega x & 0 \\
\Omega y & -1 & 0 & 0  \\ 
-\Omega x & 0 & -1 & 0 \\
0 & 0 & 0 & -1
\end{pmatrix}\, ,
\end{equation}
where $r=\sqrt{x^2+y^2}$ is the distance to the axis of rotation.  

All components of the metric tensor (\ref{eq:metric}) do not depend on time coordinate $t$. As the result, the Hamiltonian of the system which is given by the expression 
\begin{multline}\label{eq:hamiltonian}
H=\int dV \sqrt{g_{00}} \epsilon(\vec r)=\frac 1 {2g^2} \int d^3 x \sqrt {\gamma} \sqrt {g_{00}}  \biggl [ 
F^a_{t x} F^a_{t x} + F^a_{t y} F^a_{t y} + F^a_{t z} F^a_{t z} + F^a_{xy} F^a_{xy} + F^a_{xz} F^a_{xz} + F^a_{yz} F^a_{yz} - {} \\ {} -
\Omega \bigl (
x F^a_{t x} F^a_{xy} - x F^a_{t z} F^a_{yz} + y F^a_{t y} F^a_{xy} + y F^a_{t z} F^a_{xz}
\bigr ) 
\biggr ]
,
\end{multline}
is conserved. Here the Greek letters correspond to the Lorentz indices while Latin letters correspond to the color ones. The $dV = d^3 x \sqrt {\gamma}$ is the three dimensional volume in our coordinate system, $g$ is strong coupling constant and $\epsilon(\vec r)$ is energy density. Notice that to write conserved quantity it is important to multiply
the energy density $\epsilon(\vec r)$ by additional contribution of the gravitational field  $\sqrt{g_{00}}$.

Given the Hamiltonian (\ref{eq:hamiltonian}), it is straightforward to construct the partition function of the system under study
\beq
\label{eq:partition_function}
Z=\mbox{Tr}  \exp { \biggl [ - \beta  {\hat H}  \biggr ]}.
\eeq
Here one comment is in order. One can introduce the notation $T(r)=1/\beta \sqrt {g_{00}}$ which brings the expression (\ref{eq:partition_function}) to the from 
\beq
Z=\mbox{Tr}  \exp { \biggl [ - \int d V \frac {\hat \epsilon (r)} {T(r)} \biggr ]}.
\eeq
 The effective $T(r)$ is the temperature which depends on the position in space and satisfies the expression $T(r) \sqrt {g_{00}} = 1/ \beta = const$. Last equation describes  Ehrenfest–Tolman effect which states that in gravitational field the temperature is not constant in space in thermal equilibrium. 
For the rotation one has $T(r) \sqrt {1-\Omega^2 r^2}=1/\beta=T(r=0)$. So, one can conclude that rotation effectively heats the system from the rotation axis to the boundaries $T(r)>T(r=0)$. In what follows the temperature at the rotation axis $T(r=0)=1/\beta$ will be referred to as $T$.  

The calculation of the $\mbox{Tr}  \exp[...]$ in formula (\ref{eq:partition_function}) can be carried out through the standard procedure.  Applying it, the partition function of the gluodynamics in external gravitational field can be written in the following form~\cite{Yamamoto:2013zwa}
\begin{equation}\label{eq:partfunction}
Z=\int\! DA\, \exp {(-S_G)} \, .
\end{equation}
In last formula the integration over gluon degrees of freedom is carried out.  
The $S_G$ is the Euclidean action of the gluodynamics in external gravitational field, which can be written as
\begin{equation}\label{eq:ym_action}
S_{G} = \frac{1}{4 g^{2}} \int\! d^{4} x\, \sqrt{g_E}\,  g_E^{\mu \nu} g_E^{\alpha \beta} F_{\mu \alpha}^{a} F_{\nu \beta}^{a} \, .
\end{equation}
 The Euclidean metric tensor $(g_E)_{\mu\nu}$ in last formula can be obtained from (\ref{eq:metric}) by Wick rotation $t \to i \tau$. As in usual path integral for the partition function the Euclidean time $\tau$ varies in the region $\tau \in (0, \beta)$ and the gluon degrees of freedom satisfy periodic boundary conditions in temporal direction $A_{\mu}(0,{\bf x})=A_{\mu}(\beta,{\bf x})$.

Substituting the $(g_E)_{\mu\nu}$ to formula (\ref{eq:ym_action}) we get the following expression for the Euclidean action
\begin{multline}\label{eq:rot_action}
	S_{G} = \frac{1}{2 g^{2}} \int\! d^{4} x \
    \big[(1 - r^2 \Omega^2) F^a_{x y} F^a_{x y} 
    + (1 - y^2 \Omega^2) F^a_{x z} F^a_{x z} + (1 - x^2 \Omega^2) F^a_{y z} F^a_{y z} 
    + F^a_{x \tau} F^a_{x \tau} + F^a_{y \tau} F^a_{y \tau} +{} \\ {}+
    F^a_{z \tau} F^a_{z \tau} 	- 2 i y \Omega (F^a_{x y} F^a_{y \tau} + F^a_{x z} F^a_{z \tau})  
    + 2 i x \Omega (F^a_{y x} F^a_{x \tau} + F^a_{y z} F^a_{z \tau}) -
    2 x y \Omega^2 F^a_{x z} F^a_{z y}\big]\, .
\end{multline}
It is seen from this formula that the action is a complex function what leads to the sign problem. Unfortunately, direct Monte Carlo simulation of this system is impossible today. To overcome this problem instead of the real angular velocity $\Omega$, we are going to conduct Monte Carlo simulations with imaginary angular velocity $\Omega_I=-i\Omega$. The results obtained in this way will be expanded in the $\Omega_I$ and then analytically continued to real angular velocity. 

\subsection{Lattice formulation for rotating gluodynamics}\label{sec:II-lattice}

In order to conduct lattice simulation of rotating gluodynamics, one has to discretize action (\ref{eq:rot_action}). In this paper we are going use the lattice  action proposed in \cite{Yamamoto:2013zwa}, which can be written as
\begin{multline}\label{eq:rot_action_lat}
S_{G} = \frac {2 N_c} {g^2}  \sum_{x}\Big( (1 + r^{2}\Omega_I^{2}) (1 - \frac{1}{N_c} \text{Re} \Tr \ \bar{U}_{xy} ) + (1 + y^{2}\Omega_I^{2}) (1 - \frac{1}{N_c} \text{Re} \Tr \ \bar{U}_{xz} )+{} \\
{}+ (1 + x^{2}\Omega_I^{2}) (1 - \frac{1}{N_c} \text{Re} \Tr \ \bar{U}_{yz} ) +
3 - \frac{1}{N_c} \text{Re} \Tr \ 
( \bar{U}_{x \tau} + \bar{U}_{y \tau} + \bar{U}_{z \tau}) -{} \\
{}- \frac{1}{N_c} \text{Re} 
\Tr \ \big( y\Omega_I (\bar{V}_{x y \tau} + \bar{V}_{x z \tau})
- x\Omega_I ( \bar{V}_{y x \tau} +  \bar{V}_{y z \tau}) 
+ x y\Omega_I^{2}  \bar{V}_{x z y}\big) \Big),
\end{multline}
where the $\bar{U}_{\mu \nu}$ denotes clover-type average of four plaquettes (see Fig.~\ref{fig:clover}).
In the flat metric case, the use of clovers instead of plaquettes would lead to the same action after summation. However, in the case of non-uniform metric, the clover-type average allows one to build local expressions for the $F_{\mu\nu}$ and reduce discretization errors.

\begin{figure}[!h]
\begin{center}
\begin{tikzpicture}

\draw [thick, black, fill=orange, fill opacity=0.3] (45:0.2cm) rectangle (45:1.5cm);
\draw [thick, black, fill=orange, fill opacity=0.3] (135:0.2cm) rectangle (135:1.5cm);
\draw [thick, black, fill=orange, fill opacity=0.3] (-135:0.2cm) rectangle (-135:1.5cm);
\draw [thick, black, fill=orange, fill opacity=0.3] (-45:0.2cm) rectangle (-45:1.5cm);

\draw[->, thick, arrows={-latex}] (-90:1.6)--(90:1.6) node[right]{{$\mu$}};
\draw[->, thick, arrows={-latex}] (180:1.6)--(0:1.6) node[above]{$\nu$};
\filldraw[black] (-5.0,0) circle (0pt) node[anchor=west]
{\large$\overline{U}_{\mu \nu}$};
\filldraw[black] (-4.0,0,0) circle (0pt) node[anchor=west]
{\large{$=$}};
\filldraw[black] (-3.0,0,0) circle (0pt) node[anchor=west]
{\LARGE{$\frac{1}{4}$}};

\draw [decorate,decoration={brace,amplitude=10pt},xshift=-4pt,yshift=0pt]
(-1.6,-1.5) -- (-1.6, 1.5) node [black,midway,xshift=-0.6cm]{};
\draw [decorate,decoration={brace,mirror, amplitude=10pt},yshift=0pt]
(1.6,-1.5) -- (1.6, 1.5) node [black,midway,xshift=0.6cm]{};
\end{tikzpicture}
\end{center}
\caption{The clover-type average of four plaquettes.}\label{fig:clover}
\end{figure}

In order to implement the terms $F_{\mu \alpha} F_{\alpha \nu}$ on the lattice  one uses the antisymmetric chair-type
average $\bar{V}_{\mu \nu \rho}$ of 8 chairs (see Fig.~\ref{fig:chair}). The intuition for such a term in a discretized action may be understood as follows: we have terms of the form $F_{\mu \nu} F_{\nu \rho}$ in the action, and loop in the $\mu \nu$ ($\nu \rho$) plane  gives respectively $F_{\mu \nu} (F_{\nu \rho})$. The simplest gauge invariant object has to reside in $\mu \nu$ and $\nu \rho$ plane and thus involves at least 6 links.  

\begin{figure}[!h]
\begin{center}
\begin{tikzpicture}
\pgfmathsetmacro{\cubex}{1}
\pgfmathsetmacro{\cubey}{1}
\pgfmathsetmacro{\cubez}{1}
\pgfmathsetmacro{\offx}{0.1}
\pgfmathsetmacro{\offy}{0.1}
\pgfmathsetmacro{\offz}{0.1}
\pgfmathsetmacro{\distx}{5}

\draw[black, fill=cyan, fill opacity=0.15] (-\offx,-\offy,-\offz) -- ++(-\cubex,0,0) -- ++(0, 0,-\cubez) -- ++(\cubex,0,0) -- ++(0,-\cubey,0) -- ++(0,0,\cubez) -- cycle;

\draw[black, fill=cyan, fill opacity=0.15] (-\offx,-\offy, \offz) -- ++(-\cubex,0,0) -- ++(0, 0, \cubez) -- ++(\cubex,0,0) -- ++(0,-\cubey,0) -- ++(0,0,-\cubez) -- cycle;

\draw[black, fill=cyan, fill opacity=0.15] (\offx, \offy, -\offz) -- ++(\cubex,0,0) -- ++(0, 0, -\cubez) -- ++(-\cubex,0,0) -- ++(0,\cubey,0) -- ++(0,0,\cubez) -- cycle;

\draw[black, fill=cyan, fill opacity=0.15] (\offx, \offy, \offz) -- ++(\cubex,0,0) -- ++(0, 0, \cubez) -- ++(-\cubex,0,0) -- ++(0,\cubey,0) -- ++(0,0,-\cubez) -- cycle;

\draw[->, thick, arrows={-latex}] (-1.5, 0, 0)--(1.5, 0, 0) node[right]{$\mu$};
\draw[->, thick, arrows={-latex}] (0, -1.5, 0)--(0, 1.5, 0) node[left]{$\rho$};
\draw[->, thick, arrows={-latex}] (0, 0, 2)--(0, 0, -2) node[right]{$\nu$};

\draw[black] (\distx-\offx, \offy, \offz) -- ++(-\cubex,0,0) -- ++(0, 0, \cubez) -- ++(\cubex,0,0) -- ++(0,\cubey,0) -- ++(0,0,-\cubez) -- cycle;
\fill[cyan, fill opacity=0.15] (\distx-\offx, \offy, \offz) -- ++(-\cubex,0,0) -- ++(0, 0, \cubez) -- ++(\cubex,0,0) -- cycle;
\fill[cyan, fill opacity=0.15] (\distx-\offx, \offy, \offz) -- ++(0,\cubey,0) -- ++(0,0, \cubez) -- ++(0,-\cubey,0) -- cycle;
\draw[black] (\distx-\offx, \offy, -\offz) -- ++(-\cubex,0,0) -- ++(0, 0, -\cubez) -- ++(\cubex,0,0) -- ++(0,\cubey,0) -- ++(0,0,\cubez) -- cycle;
\fill[cyan, fill opacity=0.15] (\distx-\offx, \offy, -\offz) -- ++(-\cubex,0,0) -- ++(0, 0, -\cubez) -- ++(\cubex,0,0) -- ++(0,0,\cubez) -- cycle;
\fill[cyan, fill opacity=0.15] (\distx-\offx, \offy, -\offz) -- ++(0,\cubey,0) -- ++(0,0, -\cubez) -- ++(0,-\cubey,0) -- cycle;
\draw[black] (\distx+\offx, -\offy, \offz) -- ++(\cubex,0,0) -- ++(0, 0, \cubez) -- ++(-\cubex,0,0) -- ++(0,-\cubey,0) -- ++(0,0,-\cubez) -- cycle;
\fill[cyan, fill opacity=0.15] (\distx+\offx, -\offy, \offz) -- ++(\cubex,0,0) -- ++(0, 0, \cubez) -- ++(-\cubex,0,0) -- cycle;
\fill[cyan, fill opacity=0.15] (\distx+\offx, -\offy, \offz) -- ++(0,-\cubey,0) -- ++(0, 0, \cubez) -- ++(0,+\cubey,0) -- cycle;
\draw[black] (\distx+\offx, -\offy, -\offz) -- ++(\cubex,0,0) -- ++(0, 0, -\cubez) -- ++(-\cubex,0,0) -- ++(0,-\cubey,0) -- ++(0,0,\cubez) -- cycle;
\fill[cyan, fill opacity=0.15] (\distx+\offx, -\offy, -\offz)  -- ++(\cubex,0,0) -- ++(0, 0, -\cubez) -- ++(-\cubex,0,0) -- cycle;
\fill[cyan, fill opacity=0.15] (\distx+\offx, -\offy, -\offz)  -- ++(0,-\cubey,0) -- ++(0, 0, -\cubez) -- ++(0,+\cubey,0) -- cycle;

\draw[->, thick, arrows={-latex}] (-1.5 + \distx, 0, 0)--(1.5 + \distx, 0, 0) node[right]{$\mu$};
\draw[->, thick, arrows={-latex}] (0 + \distx, -1.5, 0)--(0 + \distx, 1.5, 0) node[left]{$\rho$};
\draw[->, thick, arrows={-latex}] (0 + \distx , 0, 2)--(0 + \distx, 0, -2) node[right]{$\nu$};

\filldraw[black] (-5,0,0) circle (0pt) node[anchor=west]
{\large{$\overline{V}_{\mu \nu \rho}$}};
\filldraw[black] (-3.7,0,0) circle (0pt) node[anchor=west]
{\large{$=$}};
\filldraw[black] (-3,0,0) circle (0pt) node[anchor=west]
{\LARGE{$\frac{1}{8}$}};
\draw[thick] (2.2 , 0, 0)--(2.8, 0, 0) node[right]{};

\draw [decorate,decoration={brace,amplitude=10pt},xshift=-4pt,yshift=0pt]
(-1.8,-1.8, 0) -- (-1.8, 1.8, 0) node [black,midway,xshift=-0.6cm]{};
\draw [decorate,decoration={brace,mirror, amplitude=10pt},yshift=0pt]
(7,-1.8, 0) -- (7, 1.8, 0) node [black,midway,xshift= 0.6cm]{};

\end{tikzpicture}
\end{center}
\caption{The antisymmetric chair-type average of eight chairs.}\label{fig:chair}
\end{figure}

We have performed numerical simulation on lattices $N_t\times N_z \times N_x \times N_y =N_t \times N_z \times N_s^2$ ($N_s=N_x=N_y$), such that sizes 
in spatial directions, orthogonal to the rotation axis $z$, are equal and may differ from size along the rotation axis and temporal direction. The axis $z$
is in the middle of $xy$-plane. For the all sizes considered, the restriction, following from causality: ($\Omega N_s a/\sqrt 2 < 1$) holds. And in the most cases we have $\Omega N_s a/\sqrt 2 \ll 1$. 

\subsection{Boundary conditions}

It is clear that rotating reference frame cannot be extended to arbitrary large distances from the rotation axis, since at distances $\Omega r \geq 1$ the $g_{00}$ becomes negative and such rotating system cannot be realized by real bodies. For this reason in the simulations one has to impose boundary conditions~(BC) on our system. Here we would like to stress that BC are important part of all approaches aimed at studying of rotating quark-gluon plasma rather than a lattice artifact. The results obtained within any approach depend on BC. 
In order to study this dependence in our paper we applied a series of different BC.

For all BC used in our paper we apply periodic boundary conditions for gluon fields in $z$- and $\tau$-directions. What concerns the $x$- and $y$-directions we used the following BC
\begin{enumerate}
    \item {\it Open boundary conditions (OBC).} For this type of boundary conditions the sum in action (\ref{eq:rot_action_lat}) is taken over plaquettes and chair-type products of link variables which belong to the lattice volume under investigation. If either plaquette or chair goes beyond the lattice volume it is excluded from the action. No additional restrictions are put on the link variable located on the boundary. Similar conditions were used in Ref.~\cite{Luscher:2011kk,Luscher:2014kea} for temporal direction and in Ref.~\cite{Florio:2019nte} for one of spatial directions, where topological susceptibility was investigated.\footnote{Notice that in papers \cite{Luscher:2011kk,Luscher:2014kea,Florio:2019nte} the plaquettes on the boundary were accounted in the action with the weight 1/2. We studied this variant of OBC for non-rotating lattice and did not found meaningful difference between these two approaches. } 
   
   Here we would like to mention  that OBC do not violate the $\mathbb{Z}_3$ symmetry and we do not see explicit incompatibility of OBC with the field of velocities of rotating body. For this reason, we believe that OBC are the most appropriate for the lattice simulation of rotating gluodynamics. 

    The exclusion of the plaquettes outside the studied lattice volume from the action can be considered as if one puts these plaquettes to unity what leads to zero lattice action. So, physically, OBC can be interpreted as if the studied volume is attached to classical (without quantum fluctuations) zero temperature Yang-Mills theory. Lattice study of non-rotating lattice with OBC confirms this physical picture (see Appendix~\ref{sec:app-Tc0}).
    
    \item {\it Periodic boundary conditions (PBC)} for gluon fields in $x$- and $y$-directions.  This type of BC is not compatible with the velocity distribution in the rotating body but it respects the $\mathbb{Z}_3$ symmetry of the action. 

    \item {\it Dirichlet boundary conditions (DBC).} In this case we set all links which lie on the boundary to unit matrix: $U_{\mu}(x) = 1$. The links sticking out from the lattice volume are not included in the lattice action. DBC were used in Ref.~\cite{Yamamoto:2013zwa} to study rotating QCD. In the continuum limit this corresponds to 
\begin{subequations}\label{eq:DBC}
\begin{gather}
    A_\mu (\vec{x},\tau)\vert_{x=\pm R/2} = 0, \qquad \mu=0, 2, 3, \\
    A_\mu (\vec{x},\tau)\vert_{y=\pm R/2} = 0, \qquad \mu=0, 1, 3,
\end{gather}
\end{subequations}
where $R$ is the size of  lattice in the directions $x$ and $y$. 

We do not see incompatibility of DBC with the field of velocities of rotating body, but DBC break the $\mathbb{Z}_3$ symmetry of the action. In principle, without rotation this explicit violation of the $\mathbb{Z}_3$ symmetry is finite-volume effect which disappears in the infinite volume limit (see Appendix~\ref{sec:app-Tc0}). However, as explained above, one cannot take infinite volume limit 
for the $x$ and $y$ directions in the rotating reference frame. So, in the rotating reference frame the violation of the  $\mathbb{Z}_3$ symmetry might play important role. 

It is interesting to note that physically DBC can be considered as if one fixes high temperature on the boundary of the volume under investigation. This can be understood as follows. In DBC the Polyakov loop on the boundary equals three, i.e. the $\mathbb{Z}_3$ symmetry of lattice gluodynamics is explicitly broken on the boundary. In gluodynamics breaking of $\mathbb{Z}_3$ symmetry is the property of high temperature phase. For this reason in DBC the boundary plays a role of a seed of high temperature phase or, in other words, this can be considered as if there is a  high temperature on the boundary.
From this perspective DBC is opposite to OBC with low temperature outside the lattice volume and it is particularly interesting to compare the results obtained with OBC and DBC. Numerical simulations of non-rotating lattice confirms the physical picture of high effective temperature on the boundary (see Appendix~\ref{sec:app-Tc0}).  

\end{enumerate}

One can expect that BC incompatible with the properties of rotating body might lead to unphysical behaviour of the system. Consequently, the approach applied in this paper might become inapplicable to study rotating gluodynamics. The results obtained in this paper allow us to state that for sufficiently large lattice volumes and small angular velocities BC do not change bulk properties of rotating gluodynamics considerably. We believe this fact can  be explained as follows.
For sufficiently large volume and small angular velocity it becomes energetically favorable for the system to screen BC incompatible with the bulk properties of rotating body and the bulk properties start to dominate over the screened boundary. The screening of the boundary is well seen in Fig.~\ref{fig:Lxy_open-comp} and Fig.~\ref{fig:Lxy_Dir-comp}.

\subsection{Measurement of the critical temperature}\label{sec:II-measurement}

The main question to be addressed in this paper is how rotation influences confinement/deconfinement phase transition in gluodynamics. Commonly in non-rotating gluodynamics one exploits the \emph{Polyakov loop} as an order parameter for this transition. Lattice expression for the Polyakov loop can be written as 
\begin{gather}\label{eq:polyakov_loop_local}
L(\vec x) = \text{Tr}\left[ \prod_{\tau=0}^{N_t - 1} U_4(\vec x, \tau)\right]   \, , 
\end{gather}
where $U_4(\vec x, \tau)$ is the gauge link variable in the temporal direction. 

In non-rotating gluodynamics the Polyakov loop can be used as an order parameter since lattice action of gluodynamics is invariant under the multiplication of the $U_4$ link elements in some time slice by center elements of the $SU(3)$ group ($\mathbb{Z}_3$ symmetry). At the same time the Polyakov loop is not invariant under this transformation. From the symmetry perspective, the confinement is $\mathbb{Z}_3$ symmetric phase with $\langle L \rangle =0$, whereas in the deconfinement $\mathbb{Z}_3$ symmetry is broken and $\langle L \rangle \neq 0$. If we now turn to the rotating gluodynamics, it is clear that the action (\ref{eq:rot_action_lat}) possesses $\mathbb{Z}_3$ symmetry and above arguments on treating the Polyakov loop as an order parameter persist. 

In our paper we are going to use the Polyakov loop to label the phases of rotating gluodynamics. In addition to local Polyakov loop at point $\vec x$, we are going use spatially averaged Polyakov loop
\begin{gather}
L = \frac{1}{N_s^2 N_z} \sum_{\vec{x}} L(\vec{x}) \, , \label{eq:polyakov_loop_discrete}
\end{gather}
The critical temperature of the confinement/deconfinement transition will be determined from the Polyakov loop susceptibility
\begin{equation}\label{eq:polyakov_chi}
\chi = N_s^2 N_z \left( \langle |L|^2 \rangle - \langle |L| \rangle^2 \right) \,,
\end{equation}
which has a peak at the critical temperature. We use the Gaussian fit over a set of points near the susceptibility peak to calculate the critical temperature from obtained data
\begin{equation}\label{eq:chi_fit}
    \chi(T) = A + B\, \exp{\left( -\frac{(T-T_c)^2}{2\delta T^2}\right)} \, .
\end{equation}
One might assume that due to boundary conditions and inhomogeneity of rotating gluodynamics the Polyakov loop and its susceptibility are not appropriate for finding the critical temperature. However we believe that these observables can be used to study confinement/deconfinement transition for the following reasons:
\begin{itemize}
\item In the gluodynamics with OBC (both with and without rotation) observables can depend on the coordinate, because these boundary conditions break translational symmetry. 
 However, since OBC respect $\mathbb{Z}_3$ symmetry, in the confinement phase the Polyakov loop is zero and does not depend on the spatial coordinate (see Fig. \ref{fig:Lxy_open-comp}). In the deconfinement phase the Polyakov loop is not zero and the dependence on spatial coordinate appears. 
From these facts it is clear that the spatially averaged Polyakov loop (\ref{eq:polyakov_loop_discrete}) and its susceptibility (\ref{eq:polyakov_chi}) can be used to detect the critical temperature. In Appendix~\ref{sec:app-Tc0} it is shown that in the infinite volume limit the critical temperatures in non-rotating gluodynamics with OBC and PBC agree with each other as it should be.

\item DBC also break translational symmetry. At the same time these boundary conditions, contrary to OBC and PBC, break $\mathbb{Z}_3$ symmetry even at zero temperature: the $L(\vec x)=3$ on the boundary, and, as the result, it is not zero in the bulk. Under these circumstances the Polyakov loop becomes an approximate order parameter. Still it is possible to detect the pseudo-critical temperature of this crossover through the peak of the susceptibility (\ref{eq:polyakov_chi}).  For non-rotating gluodynamics the first order phase transition and its critical temperature is recovered in the infinite volume limit $N_z \to \infty,~ N_s \to \infty$ (see Appendix~\ref{sec:app-Tc0}). Unfortunately for rotating gluodynamics the limit $N_s \to \infty$ cannot be taken and the confinement/deconfinement transition remains a crossover even in the limit $N_z \to \infty$.

\item For all BC in rotating gluodynamics the Polyakov loop acquires additional dependence on spatial coordinate due to the rotation. Because of $\mathbb{Z}_3$ symmetry, for the PBC and OBC 
 the Polyakov loop is zero and independent on the space coordinate in the confinement phase (see Fig.~\ref{fig:Lxy_periodic-comp} and Fig.~\ref{fig:Lxy_open-comp}). In the deconfinement  the Polyakov loop is non-zero and depends on the spatial coordinate.  So, the spatially averaged Polyakov loop (\ref{eq:polyakov_loop_discrete}) and its susceptibility (\ref{eq:polyakov_chi}) can be used to detect the critical temperature even in rotating gluodynamics. As was explained above, for DBC the Polyakov loop is an approximate order parameter, but its susceptibility can be used to find the pseudo-critical temperature for rotating gluodynamics. 
 
 \item The strategy used in this work is to carry out simulations at imaginary angular velocity and analytically continue the results to real angular velocity. In view of this it is important to argue that analytic continuation is reliable approach. Our study is conducted in finite lattice volume and the integration over gluon fields is taken over compact manifold. These facts allow us to state that thermodynamic functions, like, for instance, the partition function, the Polyakov loop, its susceptibility and etc.,  are  analytic functions of angular velocity in finite volume. So, the procedure of analytic continuation is justified at finite volume. If we further consider infinite volume limit $N_z \to \infty$, possible singularities in these functions might spoil analytic continuation. However, this is not the case for the critical temperature, what can be explained as follows. The confinement/deconfinement becomes the first order phase transition  in the infinite volume limit for OBC and PBC. In this case the critical temperature at finite volume $T(V)$ deviates from the critical temperature at infinite volume $T(\infty)$ by amount which scales as $T(V)-T(\infty) \sim 1/V$~\cite{Fisher:1982xt}. So, one can conduct analytic continuation at finite volume and then take infinite volume limit without facing with singularities. For DBC the simulation is even simpler. In this case the  confinement/deconfinement transition is a crossover for any $N_z$, i.e. there is no singularities and problems with analytic continuation. 

\end{itemize}

At the end of this section we would like to note that for all BC used in this paper the value of the critical temperature $T_c(\Omega_I)$ contains finite volume effects which depend on the lattice size $N_s$ (see the discussion in the Appendix~\ref{sec:app-Tc0}). In order to account for these effects below our results for the critical temperature will be presented in terms of the ratio $T(\Omega_I)/T_c(0)$. Finally, we would like to mention that the detailed description of used lattice parameters for all the BC is presented in the Appendix \ref{sec:app-list}.

\section{
The results of the calculation
}\label{sec:Tc}
\begin{figure*}[htb]
\subfigure[]{\label{fig:O-Om-pl}
\includegraphics[width=.48\textwidth]{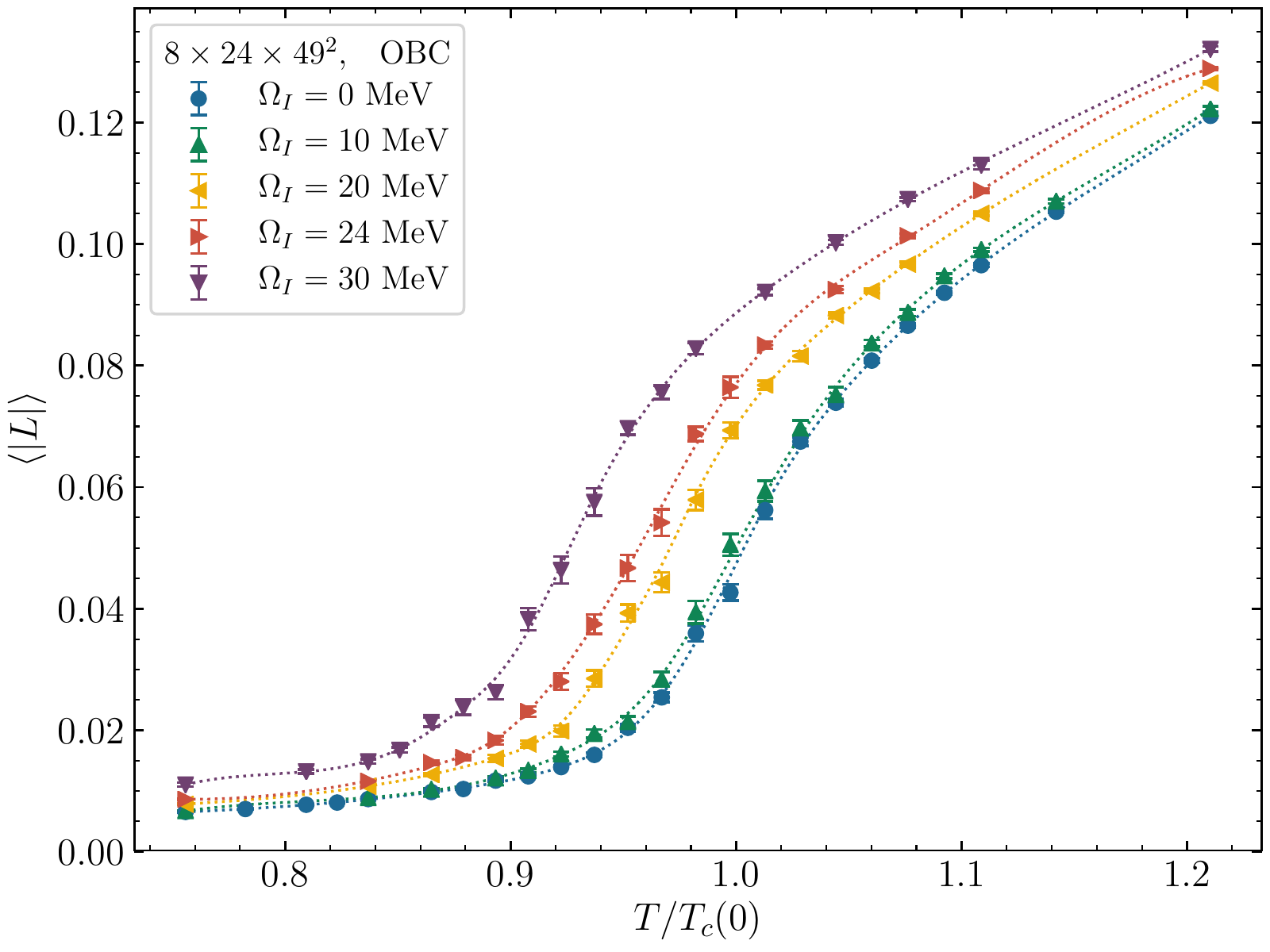}
}
\hfill
\subfigure[]{\label{fig:O-Om-chi}
\includegraphics[width=.48\textwidth]{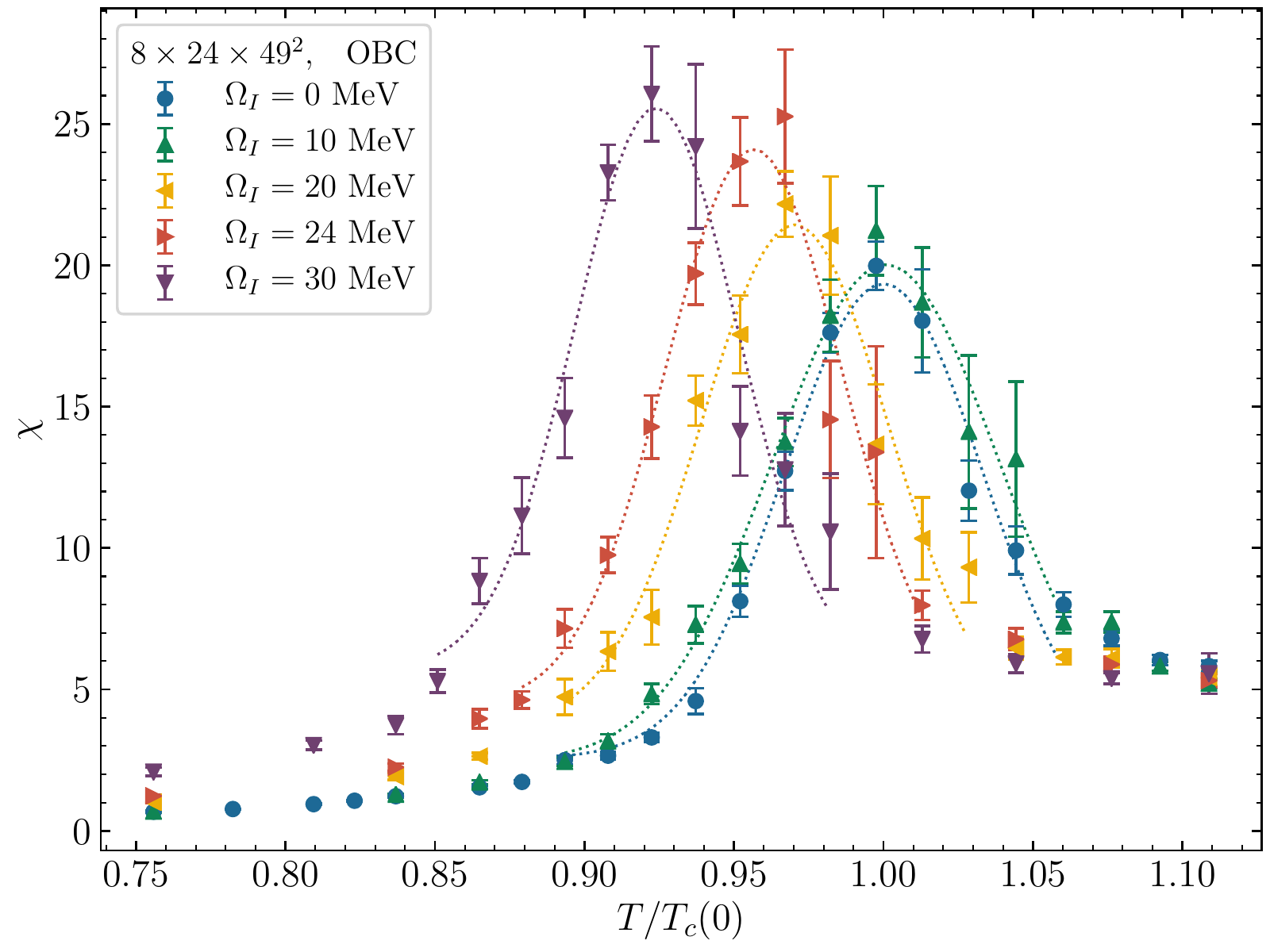}
}
\caption{The Polyakov loop~\subref{fig:O-Om-pl} and the Polyakov loop susceptibility~\subref{fig:O-Om-chi} as a function of temperature for different values of imaginary angular velocity $\Omega_I$. The results are obtained on the lattice $8\times 24\times 49^2$ with OBC. The lines for the Polyakov loop~\subref{fig:O-Om-pl} are drawn to guide the eye. The Polyakov loop susceptibilities~\subref{fig:O-Om-chi} are fitted in the vicinity of the phase transition by a Gaussian function~\eqref{eq:chi_fit}.}\label{fig:O-Om}
\end{figure*}
\subsection{Open boundary conditions}\label{sec:Tc-O}

We believe that OBC are the most appropriate for the lattice study of the rotating matter. For this reason we start the discussion of the lattice results with OBC.

\begin{figure}[htb]
\subfigure[]{\label{fig:Tc-OBC-Om}
\includegraphics[width=.48\textwidth]{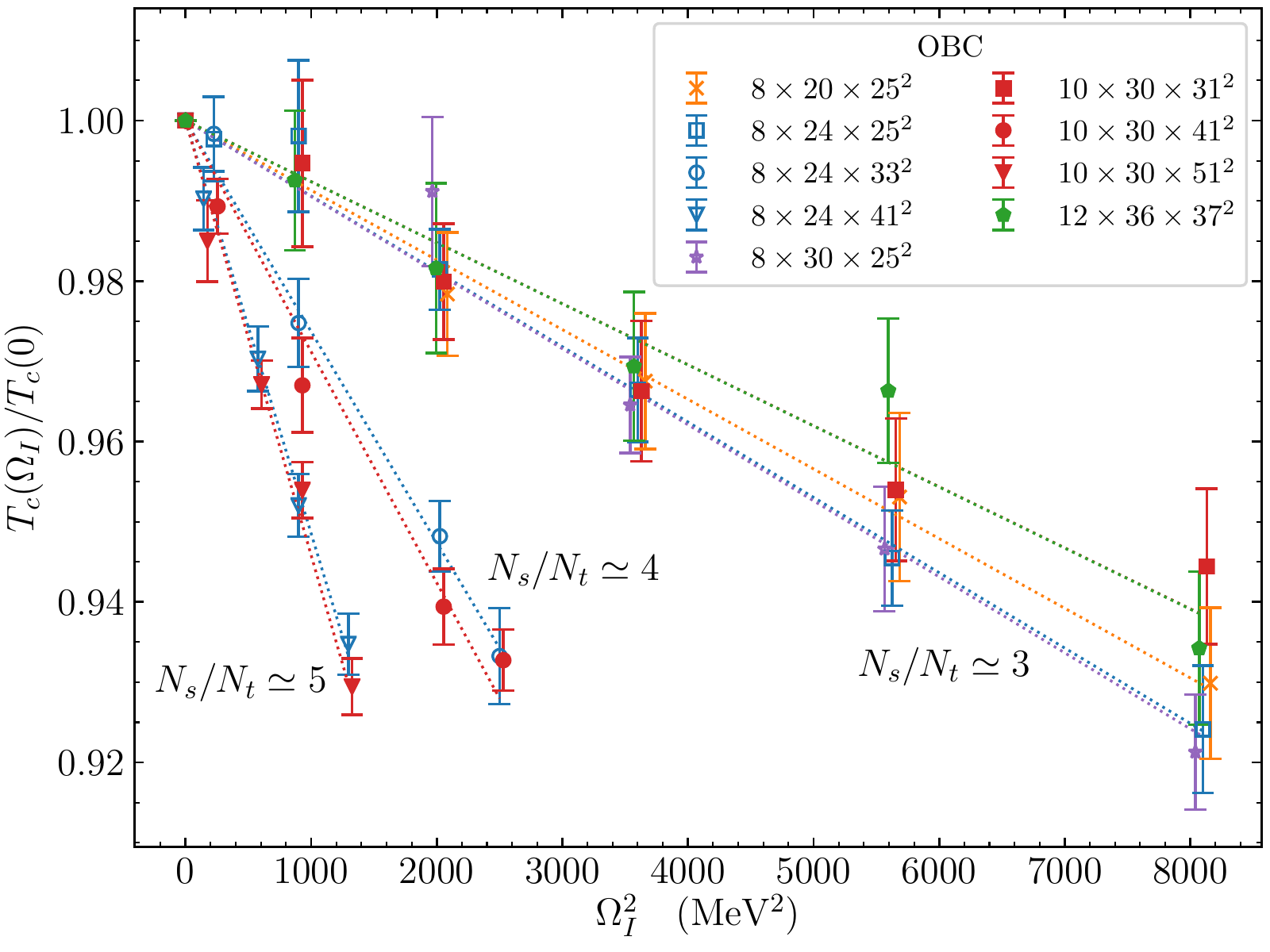}
}
\hfill
\subfigure[]{\label{fig:Tc-OBC-V}
\includegraphics[width=.48\textwidth]{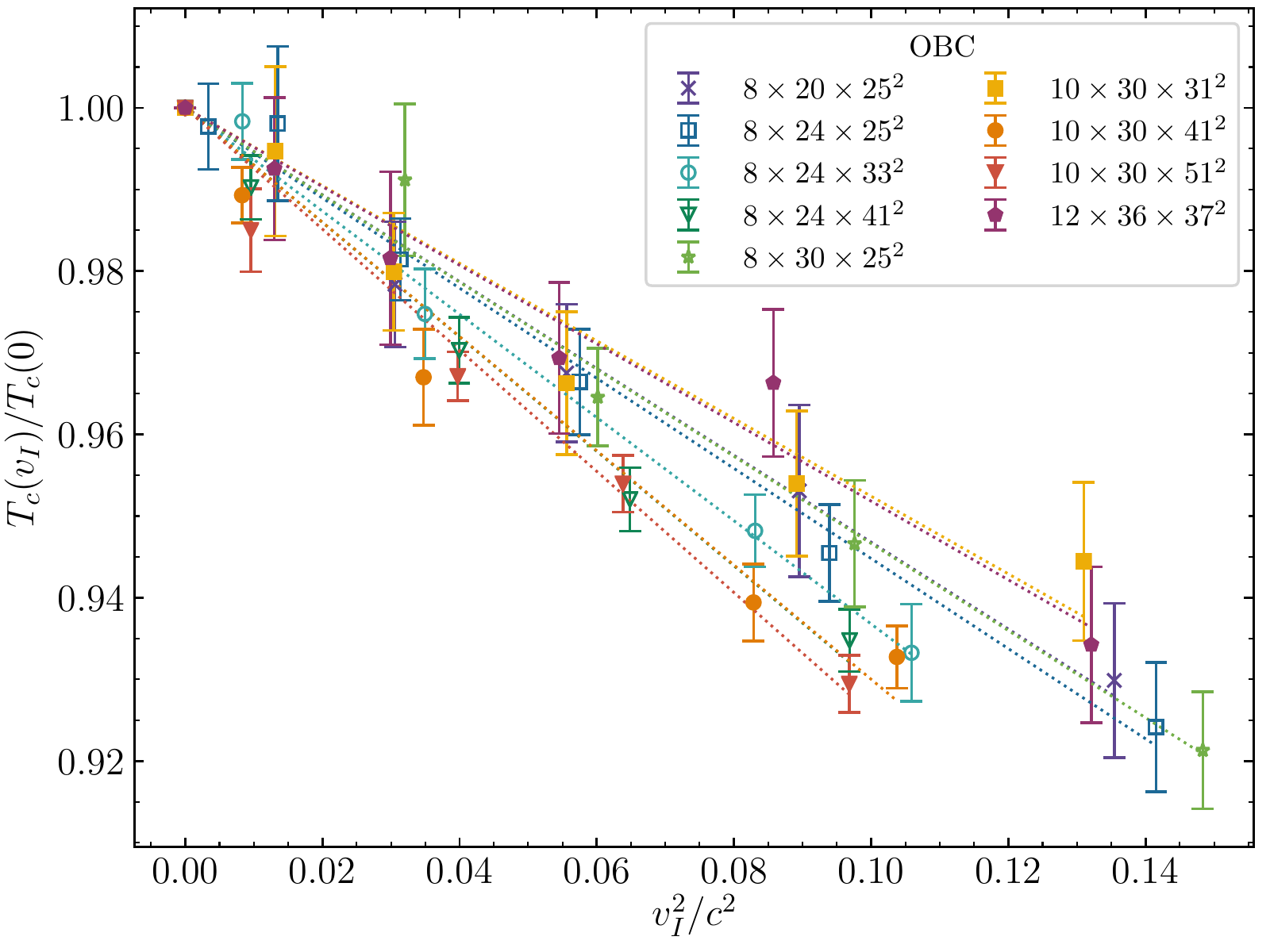}
}
\caption{The ratios $T_c/T_c(0)$ determined from the peak of the Polyakov loop susceptibility as a function of the imaginary angular velocity squared $\Omega_I^2$~\subref{fig:Tc-OBC-Om} and the linear boundary velocity squared $v_I^2$~\subref{fig:Tc-OBC-V}. Results are presented for several lattice sizes $N_t
\times N_z \times N_s^2$ with OBC. Lines correspond to simple quadratic fits $T_c(\Omega_I)/T_c(0)=1-C_2\Omega_I^2$ and $T_c(v_I)/T_c(0)=1-B_2v_I^2/c^2$}
\label{fig:Tc-OBC-var-all}
\end{figure}

The Polyakov loop and the Polyakov loop susceptibility as functions of the ratio $T/T_c(0)$ for various values of (imaginary) angular velocity $\Omega_I$ for the lattice size $8\times 24\times 49^2$ are shown in Fig.~\ref{fig:O-Om}. The confinement/deconfinement phase transition manifests itself as a rapid growth of the Polyakov loop and correspondingly as a peak in the susceptibility. One can easily read from Fig.~\ref{fig:O-Om} that the phase transition is shifted to lower temperatures when the (imaginary) angular velocity $\Omega_I$ grows. To make quantitative predictions, we fit several points in the transition region for the Polyakov loop susceptibility with Gaussian function (\ref{eq:chi_fit}). The $\chi^2/\text{ndof}$ of the fit is $\sim0.7-3$ for all angular velocities $\Omega_I$. The ratios $T_c(\Omega_I)/T_c(0)$  as functions of $\Omega_I^2$ are presented in the Fig.~\ref{fig:Tc-OBC-Om}. 
It is also instructive to introduce the (imaginary) linear velocity $v_I$ at the points with the coordinates $x=\pm{R}/{2}$, $y=0$ which are located on the boundary: $v_I = \Omega_I\, (N_s-1)\, a/2$ and to present the ratios $T_c(v_I)/T_c(0)$ as functions of $v_I$ (see Fig.~\ref{fig:Tc-OBC-V}).\footnote{Note that to determine $v_I$ we used the lattice spacing $a=a(\beta_c)$ at the critical temperature.} In order to assess systematic effects, in Fig.~\ref{fig:Tc-OBC-var-all} we present the results for various lattice sizes.

\begin{figure}[htb]
\includegraphics[width=.48\textwidth]{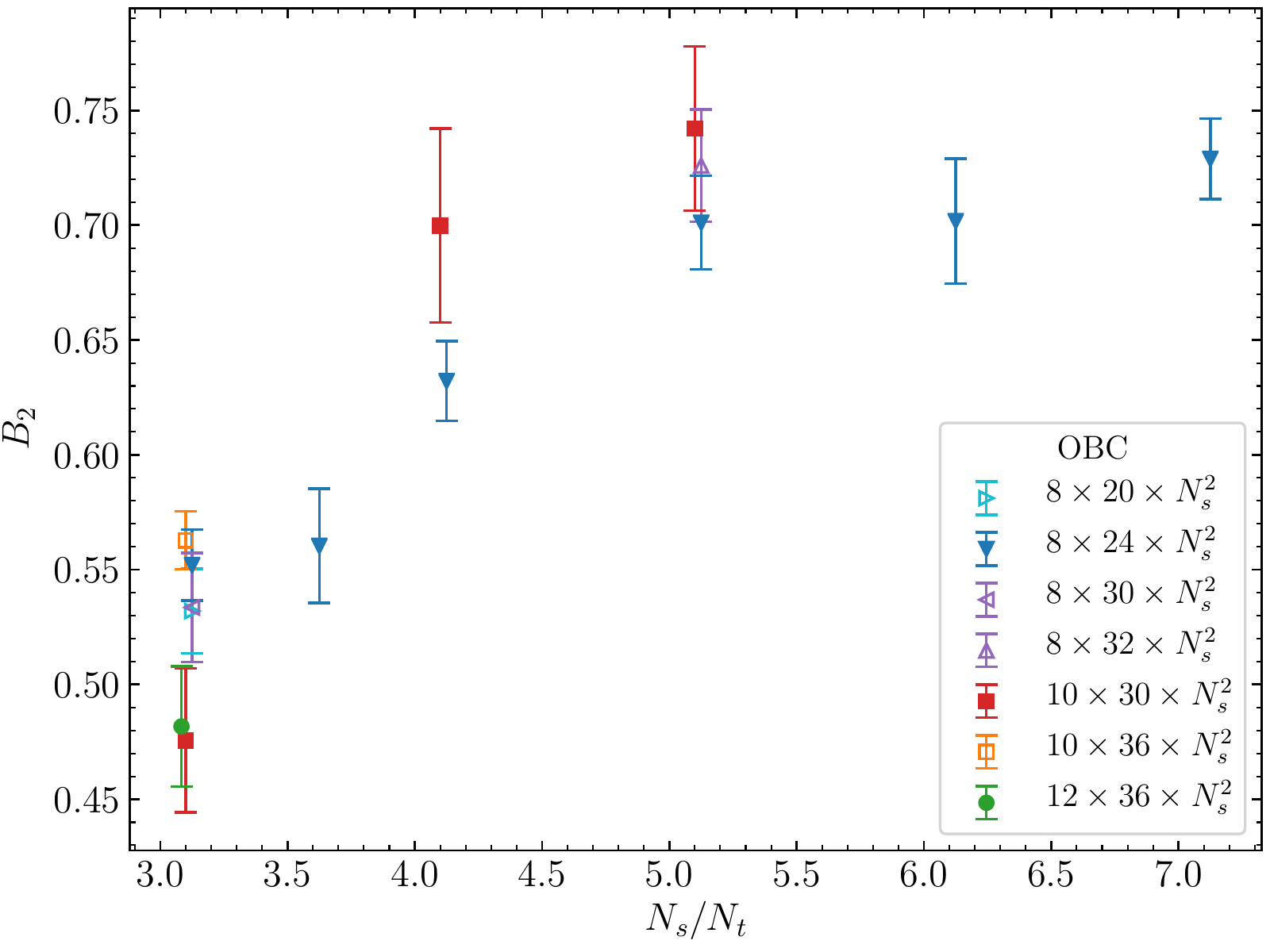}
\caption{The coefficient $B_2$ in Eq.~(\ref{eq:Tc-fitimagv}) versus the ratio $N_s/N_t$ for several lattice sizes with OBC.
}\label{fig:B2-OBC}
\end{figure}

Based on the results, presented in Fig.~\ref{fig:O-Om} and Fig.~\ref{fig:Tc-OBC-var-all}, one can draw the following conclusions:
\begin{itemize}
 \item The $T_c$ decreases with increasing $\Omega_I$. We have found that for the studied parameters the dependence of the $T_c(\Omega_I)$ on the imaginary angular velocity $\Omega_I$ can be described by a simple quadratic function ($\chi^2/\text{ndof}\sim0.4 - 2$) 
\begin{equation}\label{eq:Tc-fitimag}
    \frac{T_c(\Omega_I)}{T_c(0)} = 1 - C_2 \Omega_I^2.
\end{equation}
This confirms that the angular velocities used in the simulation are indeed small and one can expand the critical temperature in a series over $\Omega_I$.
Upon analytical continuation to real angular velocity $\Omega_I\to i\Omega$, which is legitimate for small angular velocities,  one gets the following dependence of the critical temperature on the value of the $\Omega$:
\begin{equation}\label{eq:Tc-fitreal}
    \frac{T_c(\Omega)}{T_c(0)} = 1 + C_2 \Omega^2.
\end{equation}
Our results indicate that the $C_2>0$, which leads to the conclusion: {\it With OBC the critical temperature of the confinement/deconfinement phase transition grows with increasing angular velocity.}
\begin{figure*}
\subfigure[]{\label{fig:Lxy_open-comp-0}
\includegraphics[width=0.48\textwidth]{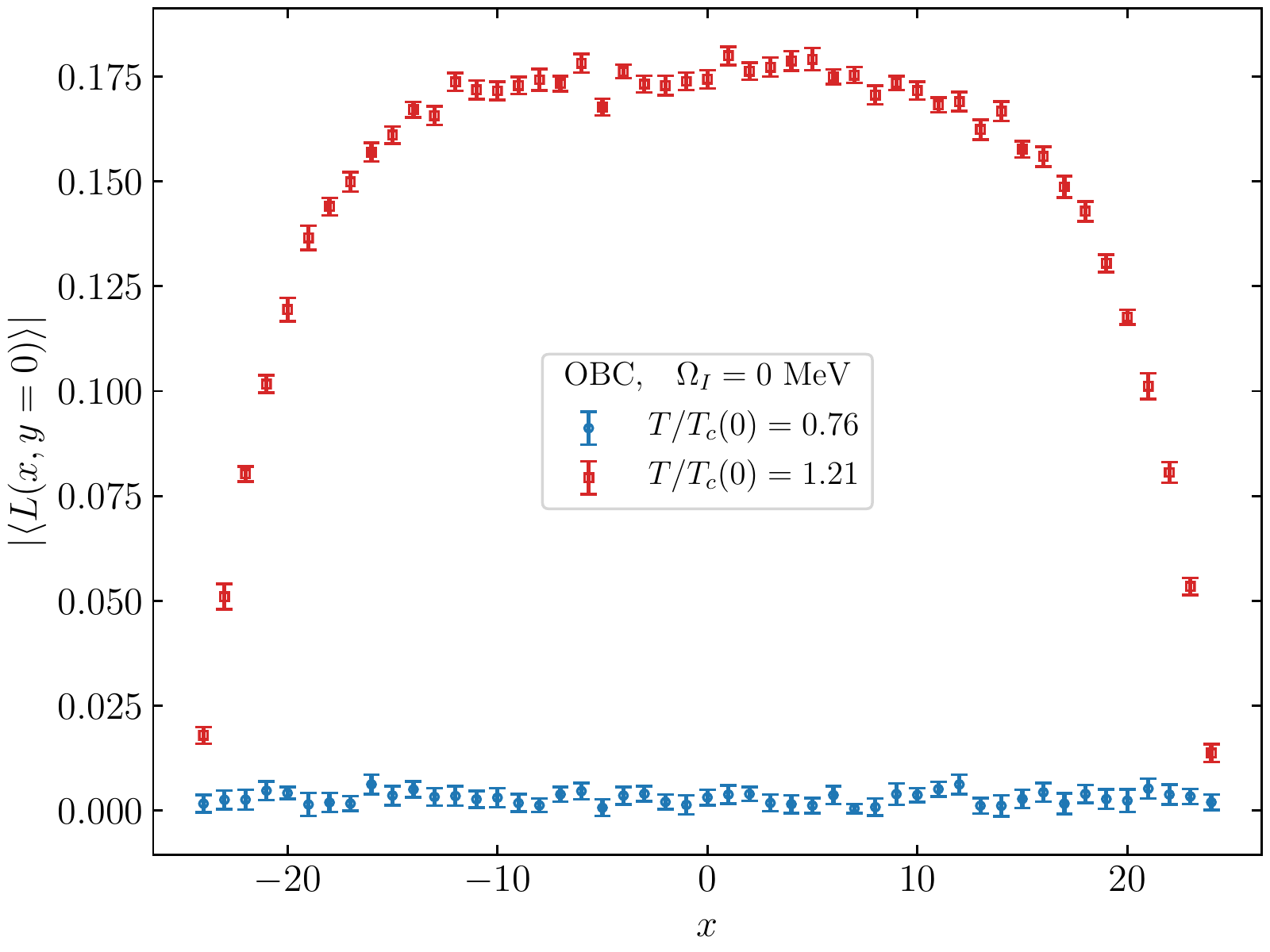}
}
\hfill
\subfigure[]{\label{fig:Lxy_open-comp-24}
\includegraphics[width=0.48\textwidth]{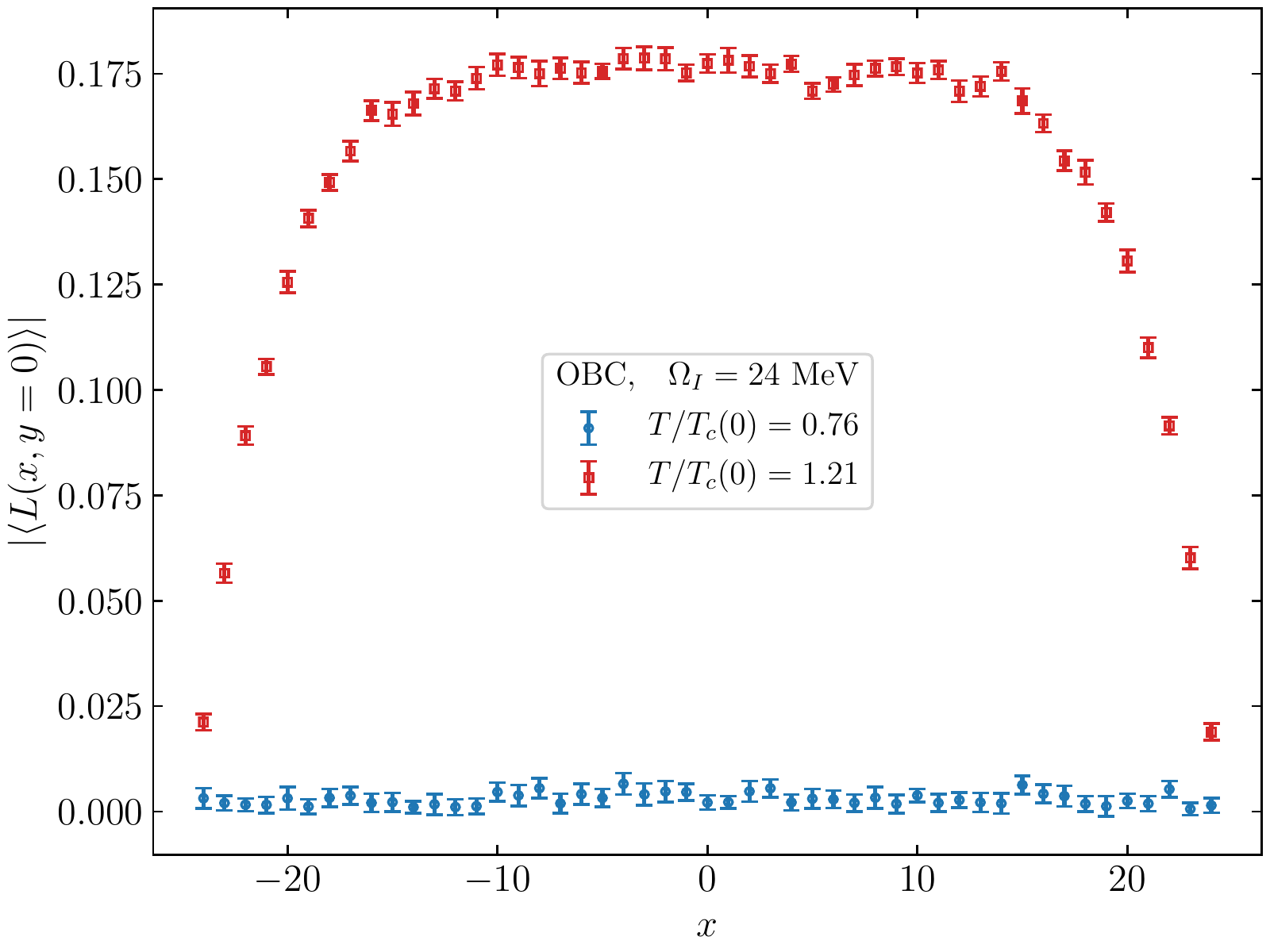}
}
\caption{The Polyakov loop $| \langle L(x,y=0) \rangle |$ as a function of coordinate $x$ for OBC and $\Omega_I = 0$ MeV \subref{fig:Lxy_open-comp-0}, $\Omega_I = 24$ MeV \subref{fig:Lxy_open-comp-24}. The results were obtained on the lattice $8\times 24 \times 49^2$ for two temperatures: $T/T_c(0)=0.76$ in the confinement phase and $T/T_c(0)=1.21$ in the deconfinement phase.}
\label{fig:Lxy_open-comp}
\end{figure*}

\item In order to study the dependence of our results on the $N_z$ lattice size we calculated the critical temperature on the lattices $8\times N_z \times 25^2,\, N_z=20,24,30$. The results obtained on these lattices agree within the uncertainty (see Fig.~\ref{fig:Tc-OBC-Om}). In order to study discretization effects, we conducted our study on the lattices $8\times 24 \times 25^2, 10\times 30 \times 31^2, 12\times 30 \times 37^2$ where the physical sizes are kept fixed. As can be seen from Fig.~\ref{fig:Tc-OBC-Om}, the ratio $T_c(\Omega_I)/T_c(0)$ shows almost no dependence on the lattice spacing $a$. Next we proceeded to the dependence of the results on 
size in the transverse directions $N_s$. To do this we fixed the $N_t$ and $N_z$ sizes and varied the $N_s$. It is seen from Fig.~\ref{fig:Tc-OBC-Om} that our data are split into lines with different slopes. The dependence of these slopes (different $C_2$ constants) on the lattice sizes $N_s$ is quite significant. 

This phenomenon can be understood in the following way.
Increasing lattice size in directions, orthogonal to the rotation axis, leads to the increase of the linear velocity on the boundary of the rotating lattice.
Since this linear velocity $\sim r\Omega$ enters the metric tensor, the action and the expressions for local temperature $T(r)=1/\beta\sqrt{g_{00}}$, it is reasonable to assume that the $T_c$ is a function of some ``velocity-like'' parameter $\sim r\Omega$, not an (imaginary) angular velocity itself. In order to check this assumption, in Fig.~\ref{fig:Tc-OBC-V} we present the ratio $T(v_I)/T_c(0)$ as a function of linear velocity $v_I^2$ on the boundary. Quite remarkably on this figure all points show a clear tendency to lie on one line. Data can be well described by a simple quadratic function
\begin{equation}\label{eq:Tc-fitimagv}
    \frac{T_c(v_I)}{T_c(0)} = 1 - B_2 \frac{v_I^2}{c^2},
\end{equation}
which corresponds to the following relation for real rotation: 
\begin{equation}\label{eq:Tc-fitrealv}
    \frac{T_c(v)}{T_c(0)} = 1 + B_2 \frac{v^2}{c^2}.
\end{equation}
We present the values of $B_2$ for several sets of parameters in Fig.~\ref{fig:B2-OBC}. The coefficient $B_2$ has a mild dependence on the parameters of the system. It does not change upon changing the lattice spacing $a$ and slightly grows with increasing lattice extent $N_s$ in $x$- and $y$-directions. It is reasonable to assume that for sufficiently large $N_s$ and small angular velocity the bulk dominates over the boundary, i.e. the role of the boundary becomes less important. We believe that this property manifests itself when the $B_2$ goes to the plateau in Fig.~\ref{fig:B2-OBC} for $N_s/N_t>4$.  So, our second conclusion is: {\it The dependence of the critical temperature on the linear velocity at the boundary $v$ has the form (\ref{eq:Tc-fitrealv}), with weak dependence of the $B_2$ on the lattice parameters. For lattices with sufficiently large $N_s$ and OBC the coefficient is $B_2\sim 0.7$.}

\item  For OBC the confinement/deconfinement  transition becomes true phase transition only in the infinite volume limit. In our lattice geometry one can only take $N_z \to \infty$ while keeping $N_s$ fixed. Since our results for the critical temperature do not depend on the $N_z$ size within the uncertainty, one concludes that the critical temperature obtained in our study is close to the infinite volume limit in the sense we have mentioned above. This also implies that the coefficients $C_2$ and $B_2$ do not depend on the $N_z$ size and they are close to the infinite volume limit. Notice that the same is true for the PBC and DBC (see Sections~\ref{sec:Tc-P} and~\ref{sec:Tc-D}). In the latter case the confinement/deconfinement transition remains a crossover even in the limit $N_z \to \infty$, but the formulas (\ref{eq:Tc-fitreal}), (\ref{eq:Tc-fitrealv}) for the crossover temperature remain to be true. The coefficients $C_2$ and $B_2$ do not depend on the $N_z$ size and they are close to the infinite volume limit as well (see Section~\ref{sec:Tc-D}).

\item It is instructive to study how the Polyakov loop depends on the spatial coordinate. Since BC and rotation break translational invariance in $x$- and $y$- directions, but preserve the invariance in $z$-direction, we introduce the local Polyakov loop in $x,y$-plane 
\begin{equation}\label{eq:polyakov_loop_local-xy}
    L(x,y) = \frac{1}{N_z} \sum_z L(x,y,z)\, ,
\end{equation}
where $L(x,y,z)=L(\vec{x})$ is defined by Eq.~\eqref{eq:polyakov_loop_local} and study its ensemble average. In Fig.~\ref{fig:Lxy_open-comp} we present the Polyakov loop $|\langle L(x, y=0)\rangle|$ as a function of the coordinate $x$ for the lattice $8\times24\times49^2$. The results are shown for two temperatures: $T/T_c(0)=0.76$ in the confinement phase and $T/T_c(0)=1.21$ in the deconfinement phase.
In addition we plot data for the lattice without rotation $\Omega_I=0$~MeV (Fig.~\ref{fig:Lxy_open-comp-0}) and with $\Omega_I=24$~MeV (Fig.~\ref{fig:Lxy_open-comp-24}). 

Fig.~\ref{fig:Lxy_open-comp} illustrates the features of Polyakov loop discussed in Section~\ref{sec:II-measurement}. One sees that Polyakov loop $|\langle L(x, y)\rangle|$ is zero for all spatial points in the confinement phase, both without rotation and with nonzero angular velocity. It confirms, that for OBC the average Polyakov loop still acts as the order parameter of confinement--deconfinement phase transition.
In the deconfinement phase one sees nontrivial coordinate dependence of Polyakov loop. Mainly this dependence can be attributed to the influence of OBC. When one moves from the boundary to the bulk, one observes that in the deconfinement phase the boundary is screened. 

\end{itemize}

\subsection{Periodic boundary conditions}\label{sec:Tc-P}

\begin{figure*}[htb]
\subfigure[]{\label{fig:P-Om-pl}
\includegraphics[width=.48\textwidth]{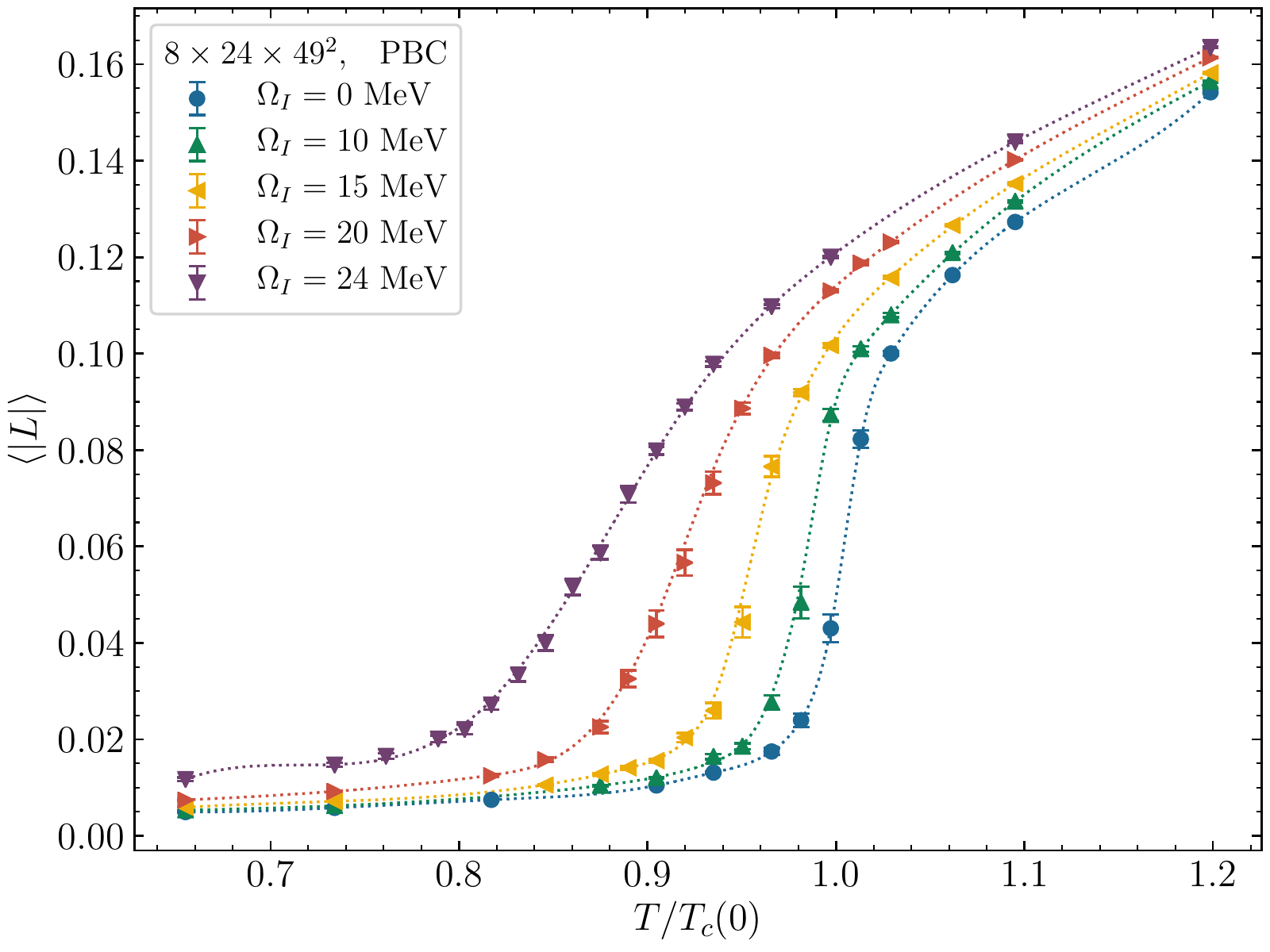}
}
\hfill
\subfigure[]{\label{fig:P-Om-chi}
\includegraphics[width=.48\textwidth]{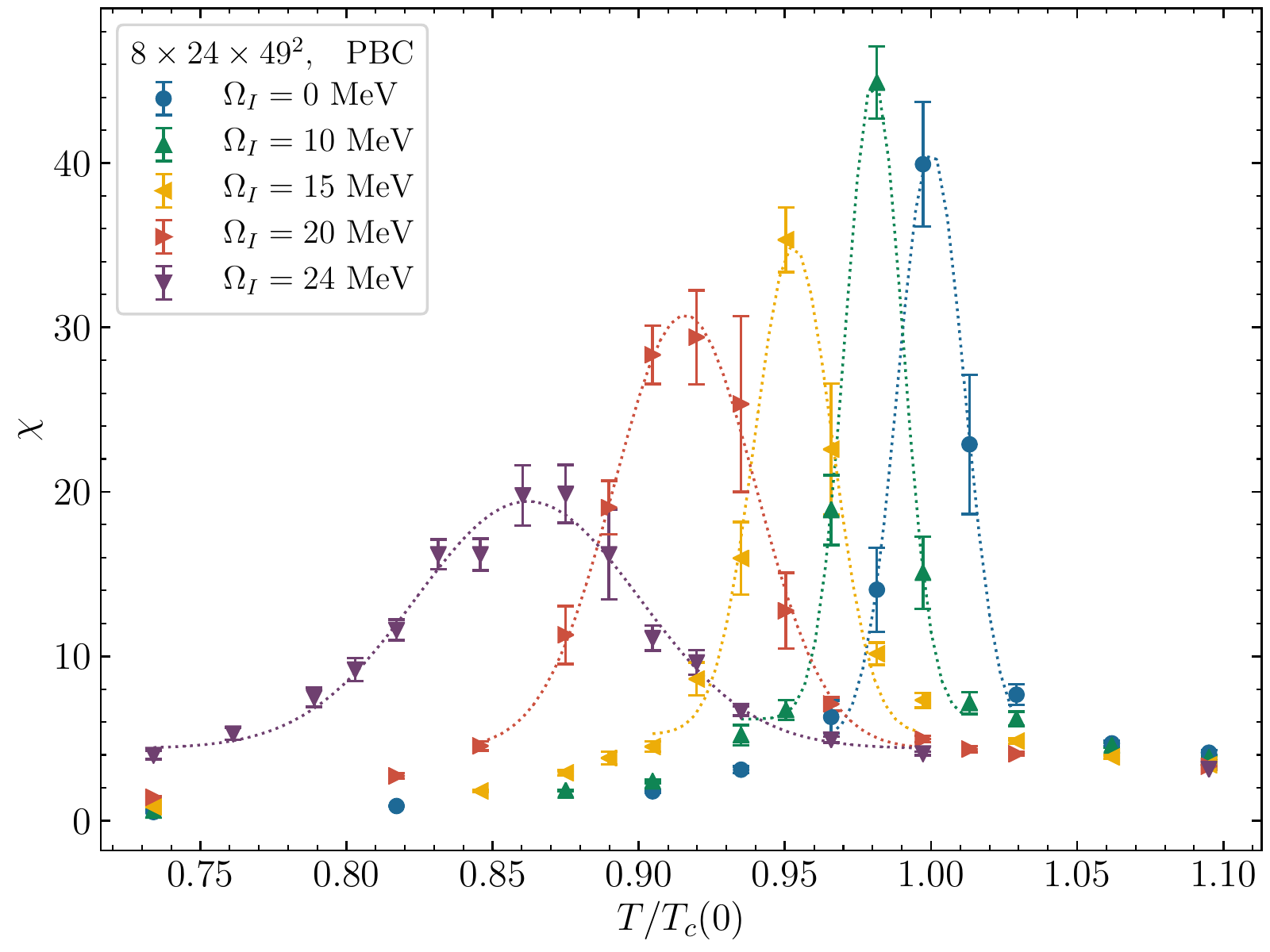}
}
\caption{The Polyakov loop~\subref{fig:P-Om-pl} and the Polyakov loop susceptibility~\subref{fig:P-Om-chi} as a function of temperature for different values of imaginary angular velocity $\Omega_I$. The results are obtained on the lattice $8\times 24\times 49^2$ with PBC. The lines for the Polyakov loop~\subref{fig:P-Om-pl} are drawn to guide the eye. The Polyakov loop susceptibilities~\subref{fig:P-Om-chi} are fitted in the vicinity of the phase transition by the Gaussian function~\eqref{eq:chi_fit}.
}\label{fig:P-Om}
\end{figure*}
In this section we present the results of our study of the confinement/deconfinement phase transition for rotating gluodynamics with PBC. Although PBC are commonly used in lattice simulations, they might not be physical for the simulation of rotating medium. Nonetheless, we believe that they can be used to check the robustness of our predictions against changing the boundary conditions.   

In Fig.~\ref{fig:P-Om} we present the Polyakov loop and the Polyakov loop susceptibility with respect to the temperature for various values of (imaginary) angular velocity $\Omega_I$ for the lattice size $8\times 24\times 49^2$. 
It is clearly seen that the behaviour of the critical temperature  is the same as for OBC: it decreases with growing imaginary angular velocity.

To determine the critical temperature we fitted the susceptibility in the vicinity of the peak by the Gaussian function~(\ref{eq:chi_fit}). In Fig.~\ref{fig:Tc-PBC-var-ls} we present the obtained dependence of the ratios $T_c/T_c(0)$ on the imaginary angular velocity $
\Omega_I$ (Fig.~\ref{fig:Tc-PBC-Om}) and the corresponding linear boundary velocity $v_I$ (Fig.~\ref{fig:Tc-PBC-V}) for several lattice sizes.

In general, the behaviour of the confinement/deconfinement phase transition in the rotating gluodynamics with PBC is very similar to the case with OBC. In particular, it is worth to mention the following:

\begin{figure}[htb]
\subfigure[]{\label{fig:Tc-PBC-Om}
\includegraphics[width=.48\textwidth]{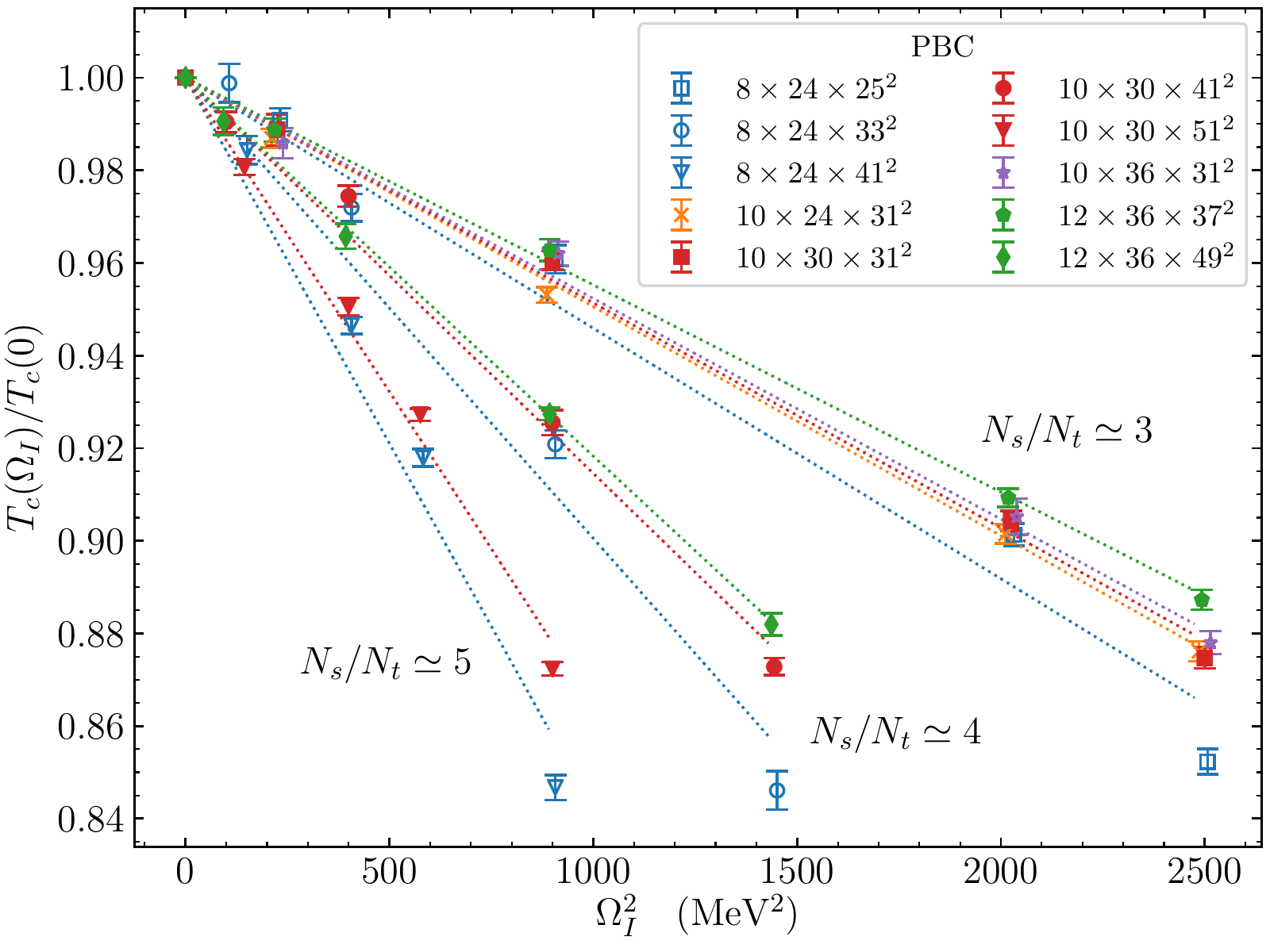}
}
\hfill
\subfigure[]{\label{fig:Tc-PBC-V}
\includegraphics[width=.48\textwidth]{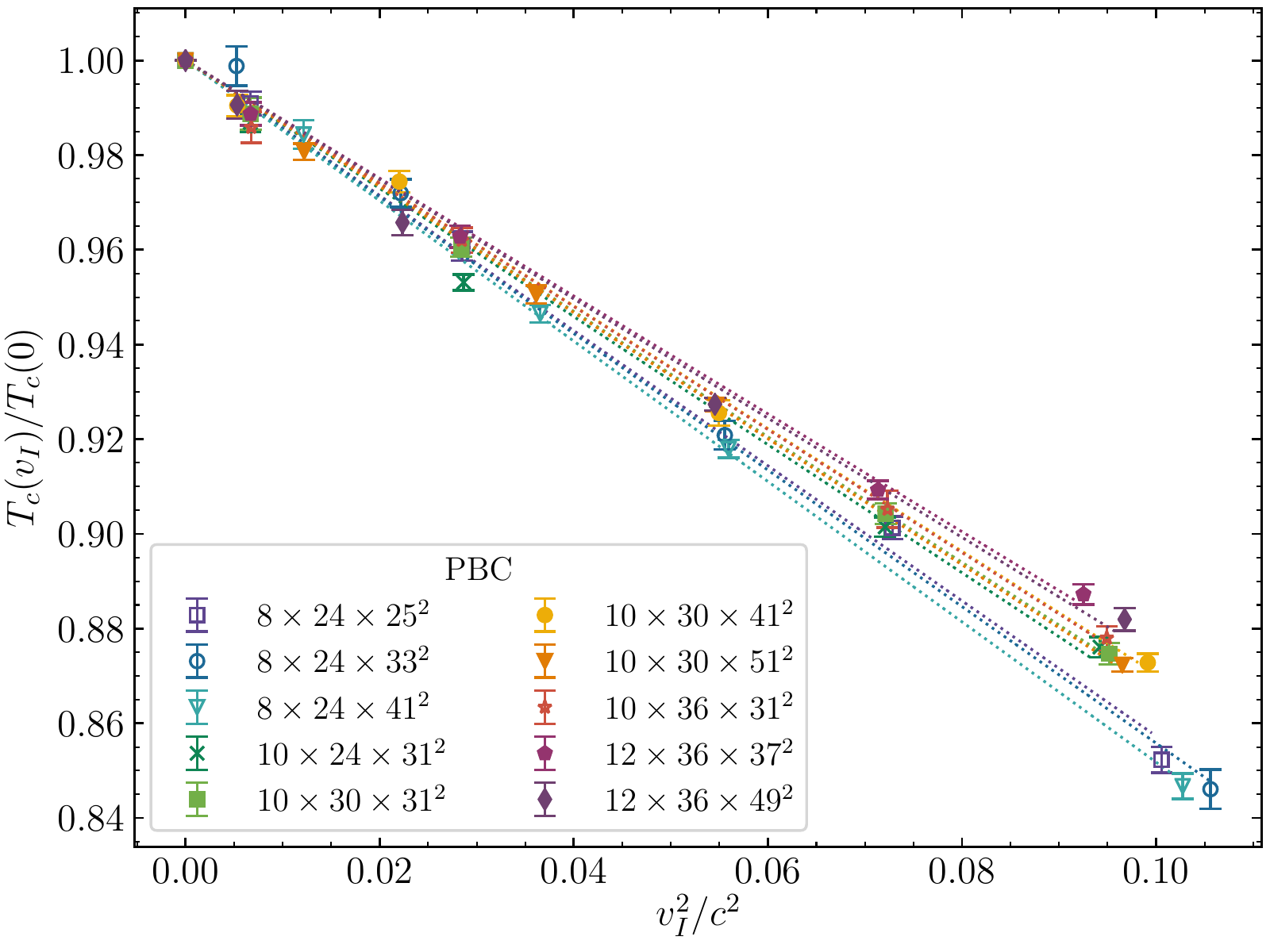}
}
\caption{
The ratios $T_c/T_c(0)$ determined from the peak of the Polyakov loop susceptibility as a function of the imaginary angular velocity squared $\Omega_I^2$~\subref{fig:Tc-PBC-Om} and the linear boundary velocity squared $v_I^2$~\subref{fig:Tc-PBC-V}. Results are presented for several lattice sizes $N_t
\times N_z \times N_s^2$ with PBC. Lines correspond to simple quadratic fits $T_c(\Omega_I)/T_c(0)=1-C_2\Omega_I^2$ and $T_c(v_I)/T_c(0)=1-B_2v_I^2/c^2$
}\label{fig:Tc-PBC-var-ls} 
\end{figure}

\begin{itemize}

    \item In Fig.~\ref{fig:Tc-PBC-Om} we present the dependence of the ratio $T_c(\Omega_I)/T_c(0)$ on the imaginary angular velocity $\Omega_I$. One can easily see that  
    with good accuracy it can be described by the same formula, as for OBC: $T_c(\Omega_I)/T_c(0)=1-C_2\Omega_I^2,$ with $C_2>0$. After analytical continuation we draw a conclusion that {\it with PBC the critical temperature of the confinement/deconfinement transition increases with angular velocity}.
    
    \item Analogously to OBC, with PBC we have performed simulations for several lattice sizes, in order to estimate systematic uncertainties. It is seen that uncertainties of the calculation for PBC are smaller than that for OBC.
    
    By looking at the results of the simulations with lattice sizes $10\times N_z\times31^2$, $N_z=24,30,36$ we conclude that effects of finite lattice size in $z$-direction are small. From the results obtained on the lattice sizes $8\times24\times25^2$, $10\times30\times31^2$ and $12\times36\times37^2$ with fixed physical volumes one can read off, that the dependence on the lattice spacing is rather weak. However, it can be easily seen that when one varies lattice size $N_s$, the ratio $T_c(\Omega_I)/T_c(0)$ changes significantly. Again, similarly to OBC, this dependence can be absorbed by looking at the $T_c(v_I)/T_c(0)$ versus the linear boundary velocity squared $v_I^2$ (see Fig.~\ref{fig:Tc-PBC-V}): $T_c(v_I)/T_c(0)=1-B_2v_I^2/c^2$. 
    
    We present the results for the coefficient $B_2$ in Fig.~\ref{fig:B2-PBC}. It is worth noting, that the values of the coefficient $B_2$ for $N_t=8$ are slightly larger, then for $N_t=10$ and $12$, which almost do not differ with each other within errorbars. It may be attributed to finite lattice spacing effects. For the $N_t=8$ lattices the dependence of the $B_2$ on $N_s$ is either very slowly rising with $N_s$ or constant within the uncertainty. While for the $N_t=10,12$ lattices there is no such dependence within the uncertainty of the calculation. We thus conclude, that {\it for PBC the relation between the critical temperature and the linear boundary velocity also has the form (\ref{eq:Tc-fitrealv})} with $B_2\sim 1.3$.

\begin{figure}[tb]
\includegraphics[width=.48\textwidth]{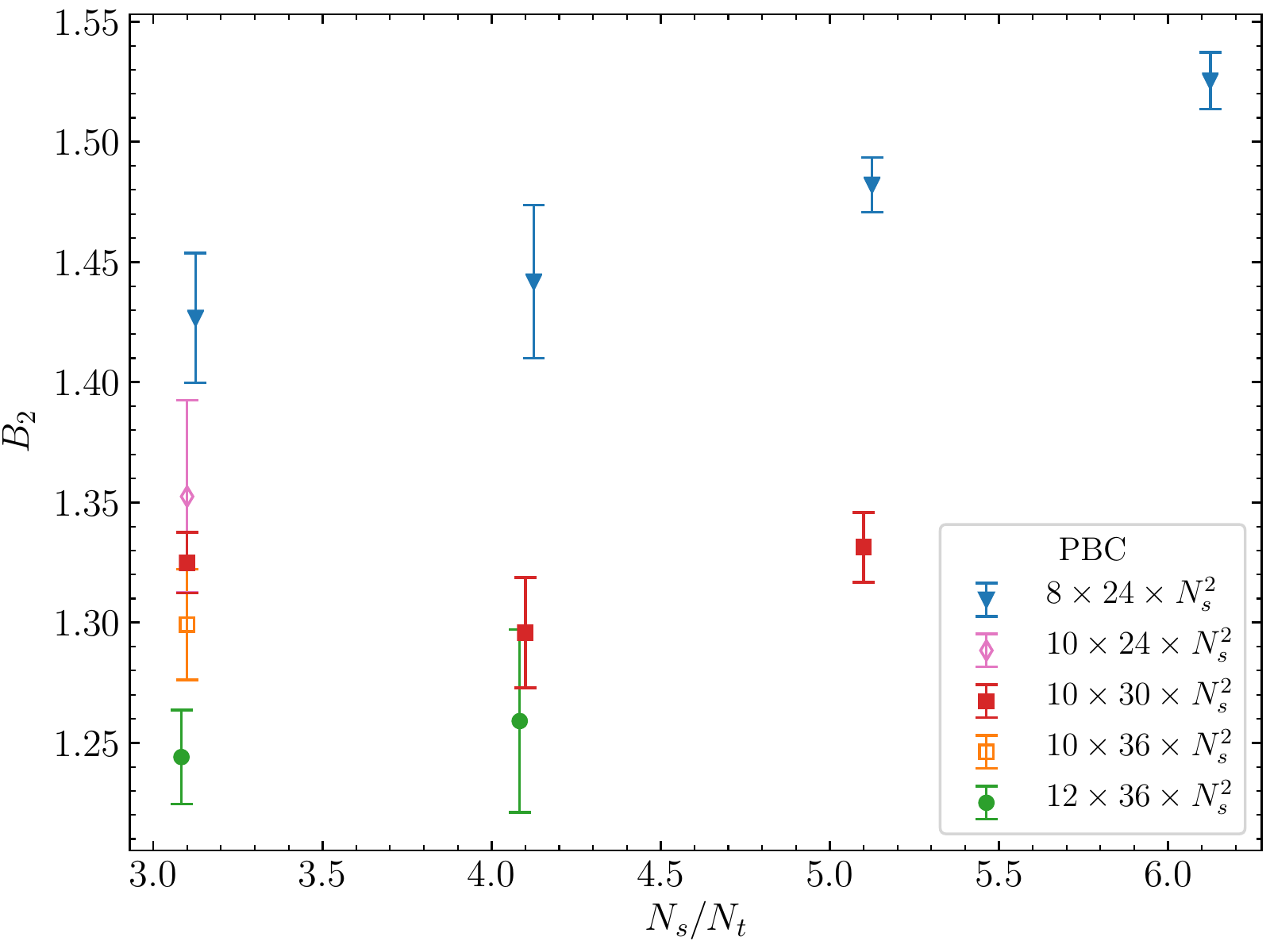}
\caption{
The coefficient $B_2$ in Eq.~(\ref{eq:Tc-fitimagv}) versus the ratio $N_s/N_t$ for several lattice sizes with PBC.
}\label{fig:B2-PBC} 
\end{figure}

\begin{figure*}
\subfigure[]{\label{fig:Lxy_periodic-comp-0}
\includegraphics[width=0.48\textwidth]{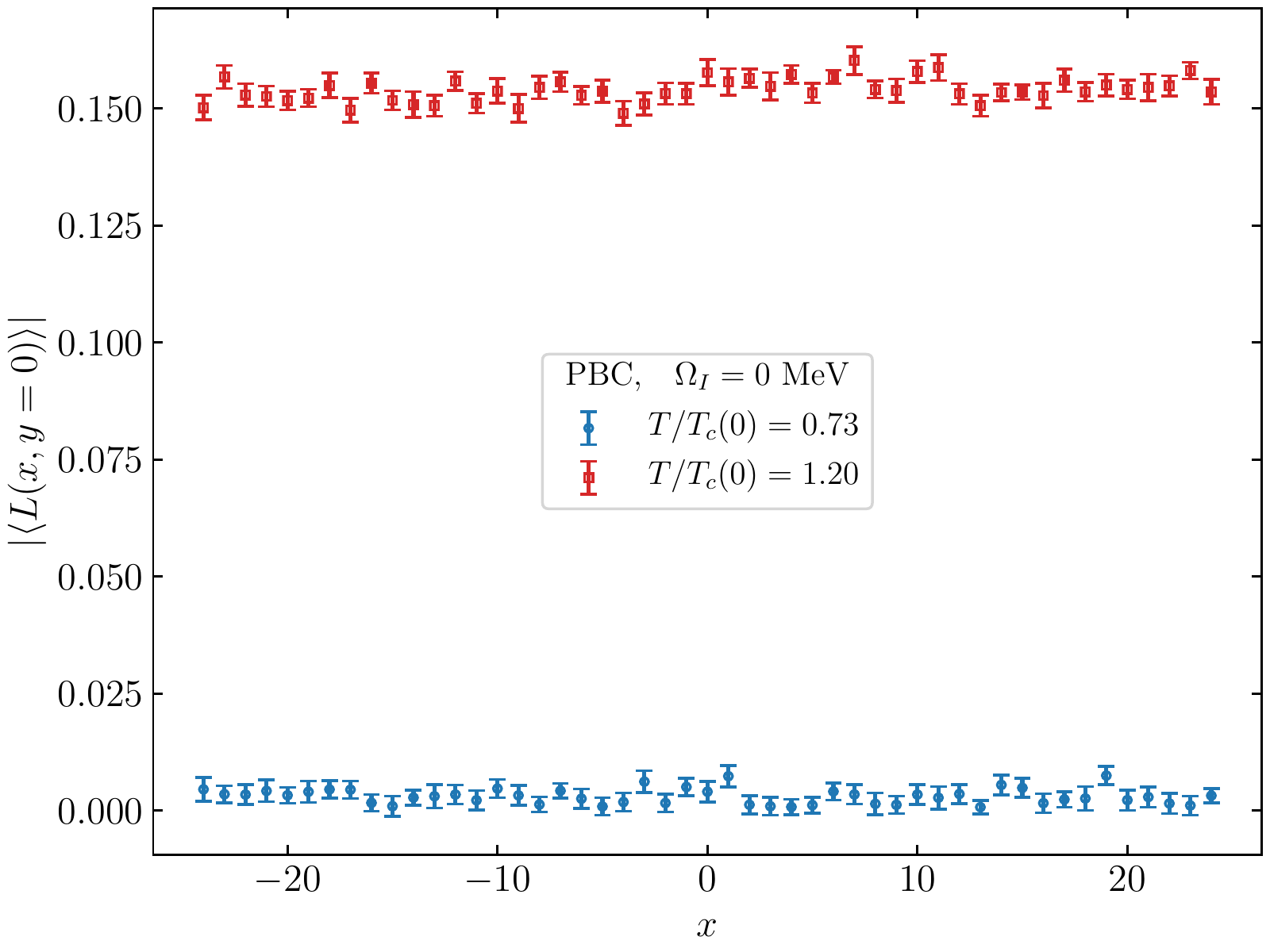}
}
\hfill
\subfigure[]{\label{fig:Lxy_periodic-comp-24}
\includegraphics[width=0.48\textwidth]{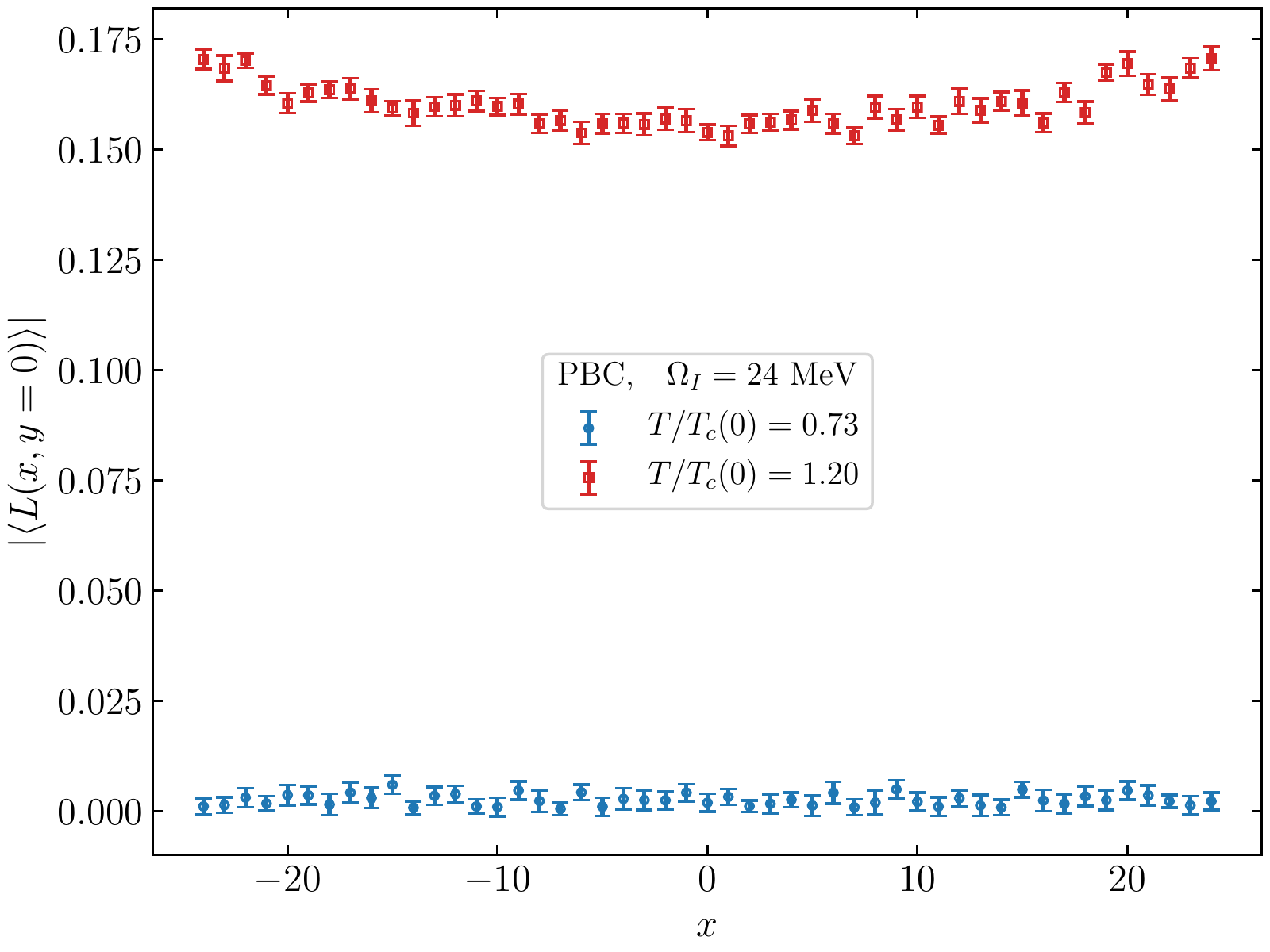}
}
\caption{The Polyakov loop $| \langle L(x,y=0) \rangle |$ as a function of coordinate $x$ for PBC and $\Omega_I = 0$ MeV \subref{fig:Lxy_open-comp-0}, $\Omega_I = 24$ MeV \subref{fig:Lxy_open-comp-24}. The results were obtained on the lattice $8\times 24 \times 49^2$ for two temperatures: $T/T_c(0)=0.73$ in the confinement phase and $T/T_c(0)=1.20$ in the deconfinement phase.}
\label{fig:Lxy_periodic-comp}
\end{figure*}
    \item  Similarly to Fig.~\ref{fig:Lxy_open-comp}, in Fig.~\ref{fig:Lxy_periodic-comp} we present the Polyakov loop $|\langle L(x, y=0)\rangle|$ as a function of the coordinate $x$ for the lattice $8\times24\times49^2$ with PBC. The results are shown for two temperatures: $T/T_c(0)=0.73$ in the confinement phase and $T/T_c(0)=1.20$ in the deconfinement phase.
In addition we plot data for the lattice without rotation $\Omega_I=0$~MeV (Fig.~\ref{fig:Lxy_periodic-comp-0}) and with $\Omega_I=24$~MeV (Fig.~\ref{fig:Lxy_periodic-comp-24}). With and without rotation Polyakov loop is zero in the confinement phase, whereas it  develops nonzero values in the deconfinement phase. Comparing Fig.~\ref{fig:Lxy_periodic-comp-0} and Fig.~\ref{fig:Lxy_periodic-comp-24} it is seen that the Polyakov loop acquires weak dependence on the coordinate due to the rotation. 

\end{itemize}

\subsection{Dirichlet boundary conditions}\label{sec:Tc-D}

\begin{figure*}[htb]
\subfigure[]{\label{fig:D-Om-pl}
\includegraphics[width=.48\textwidth]{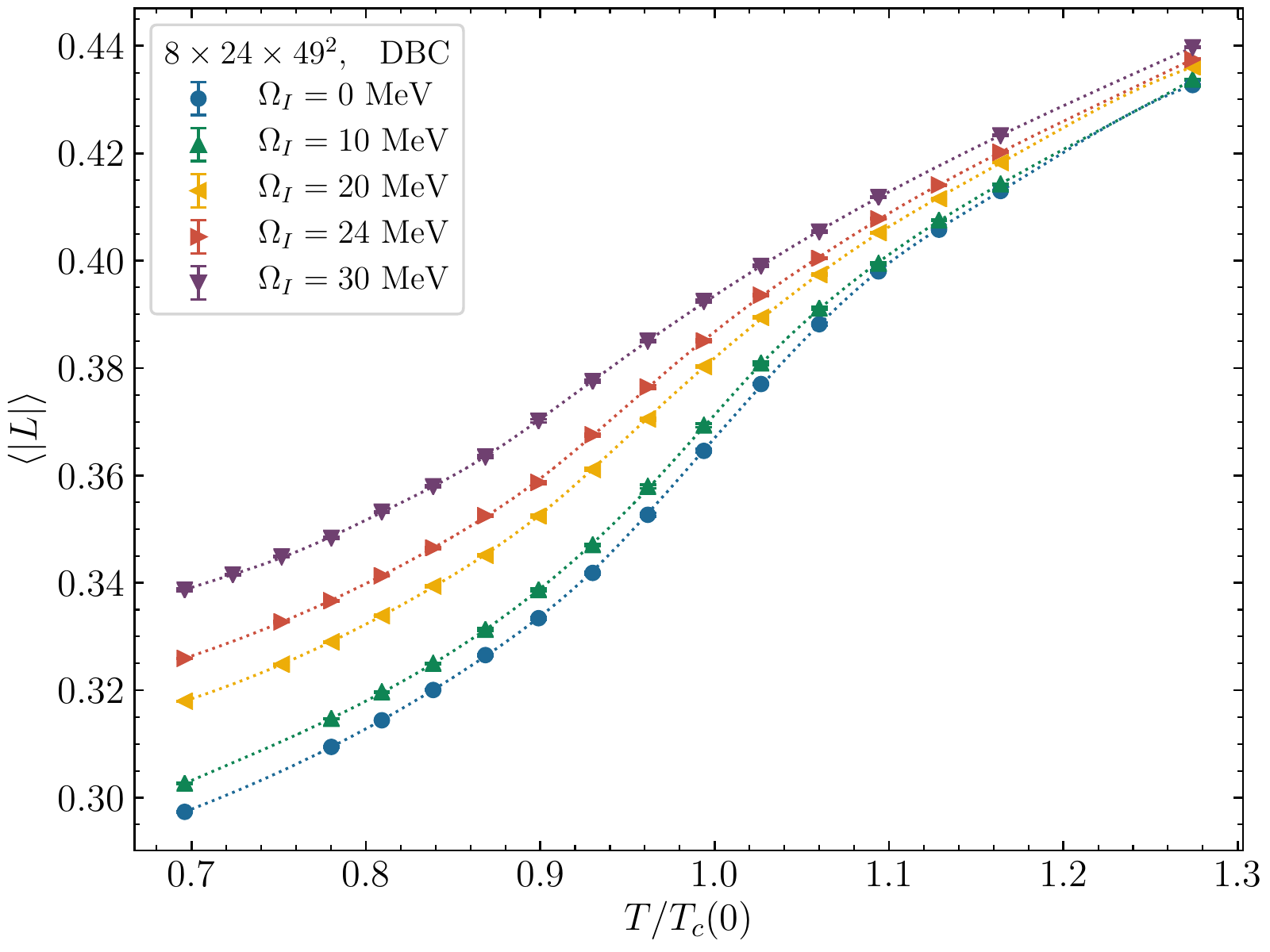}
}
\hfill
\subfigure[]{\label{fig:D-Om-chi}
\includegraphics[width=.48\textwidth]{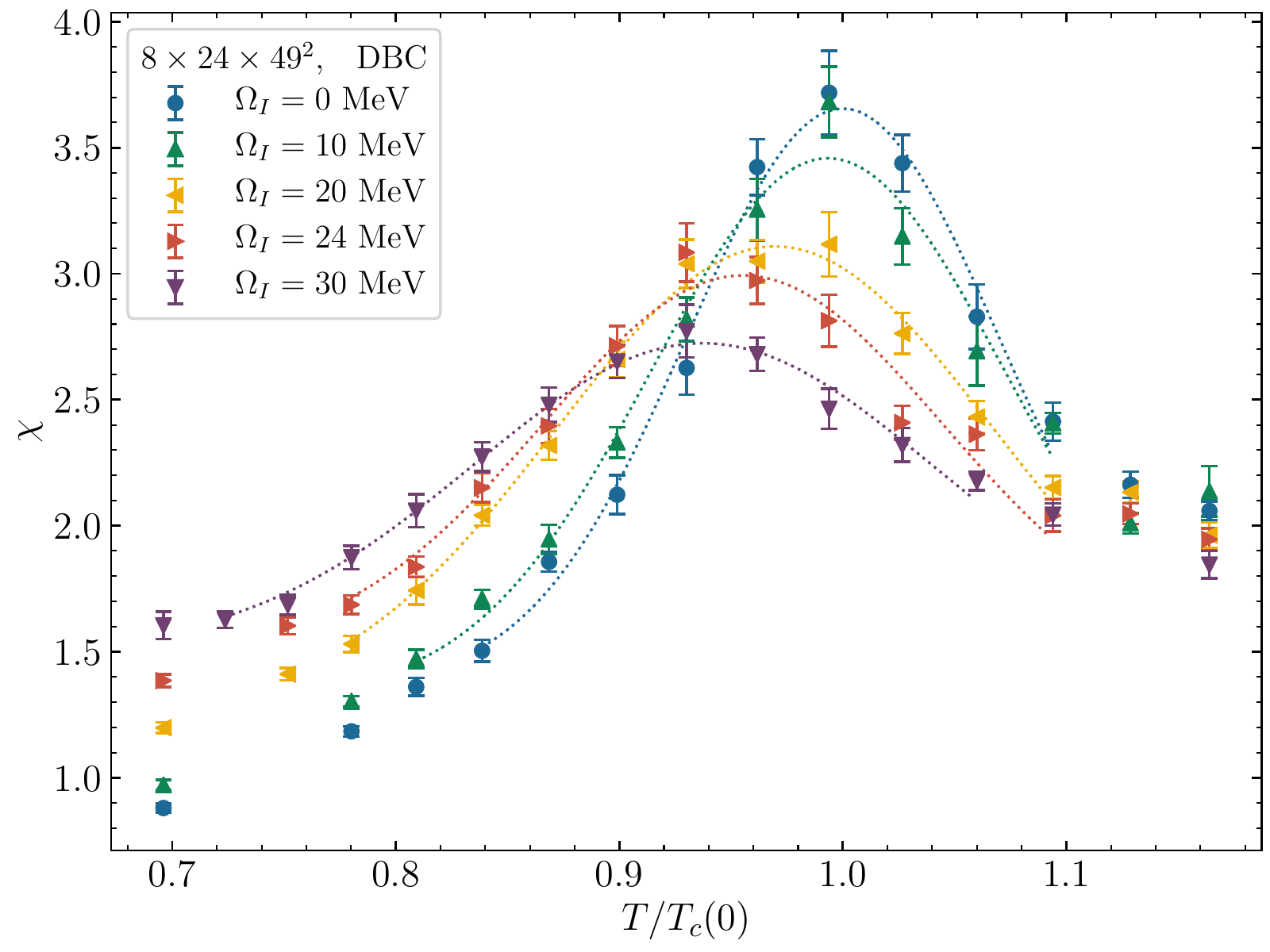}
}

\caption{The Polyakov loop~\subref{fig:D-Om-pl} and the Polyakov loop susceptibility~\subref{fig:D-Om-chi} as a function of temperature for different values of imaginary angular velocity $\Omega_I$. The results are obtained on the lattice $8\times 24\times 49^2$ with DBC. The lines for the Polyakov loop~\subref{fig:D-Om-pl} are drawn to guide the eye. The Polyakov loop susceptibilities~\subref{fig:D-Om-chi} are fitted in the vicinity of the phase transition by the Gaussian function~\eqref{eq:chi_fit}. }\label{fig:D-Om}
\end{figure*}

In this section we present the results  for the phase diagram of the rotating gluodynamics with DBC. These boundary conditions explicitly break $\mathbb{Z}_3$ symmetry what leads to additional lattice artifacts, which disappear only in the thermodynamic limit (see Appendix~\ref{sec:app-Tc0}). As a consequence of the explicit center symmetry breaking, it is natural to expect that they make the phase transition smoother. In order observe good peak in the susceptibility of the Polyakov loop one has to conduct lattice simulations on the lattices with much larger spatial volumes as compared to OBC and DBC. 
For this reason DBC are more expensive from computational point of view and we made an exploratory study of these BC with few investigated lattice sizes.

\begin{figure}[htb]
\subfigure[]{\label{fig:Tc-DBC-Om}
\includegraphics[width=.48\textwidth]{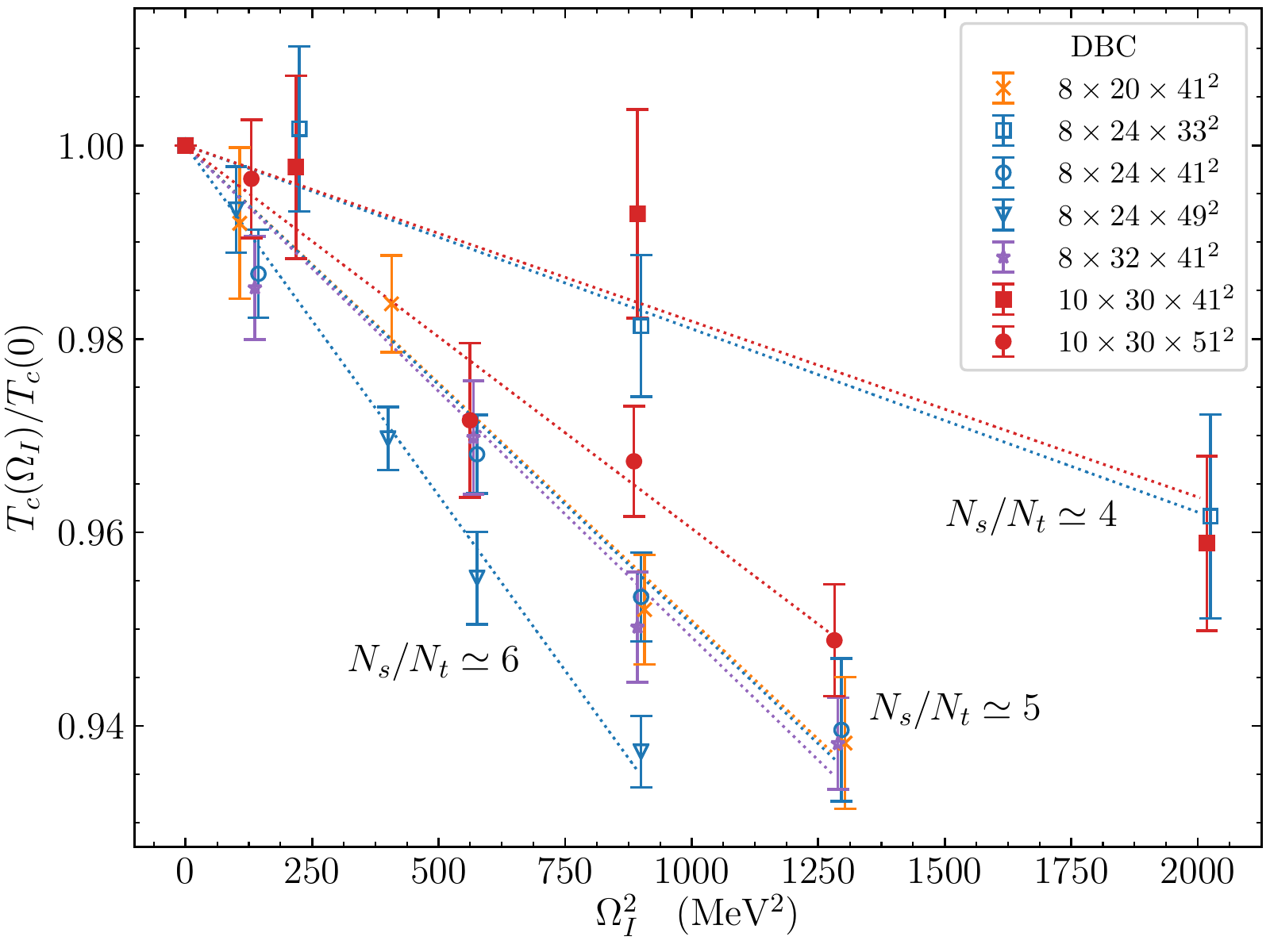}
}
\hfill
\subfigure[]{\label{fig:Tc-DBC-V}
\includegraphics[width=.48\textwidth]{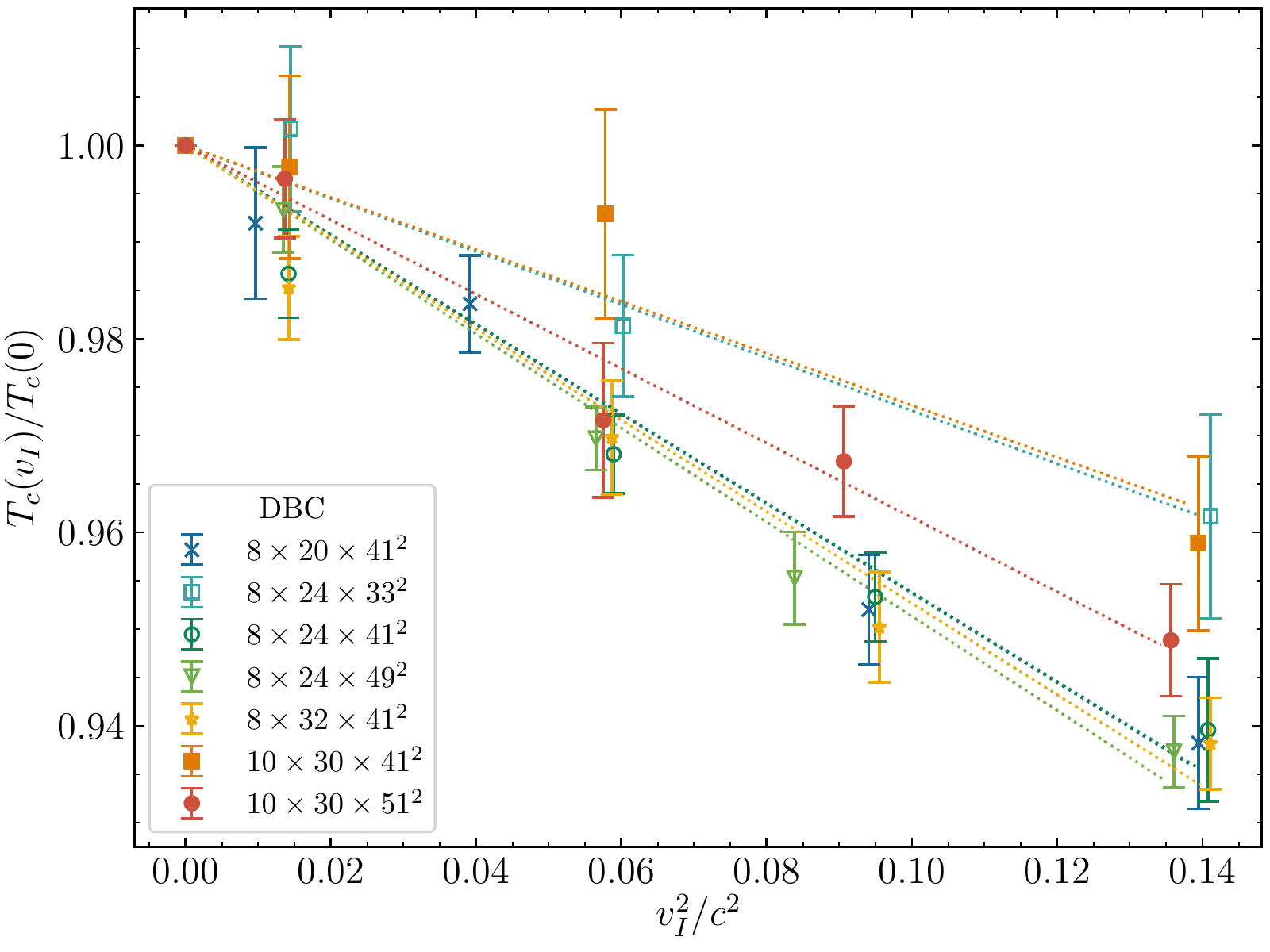}
}
\caption{
The ratios $T_c/T_c(0)$ determined from the peak of the Polyakov loop susceptibility as a function of the imaginary angular velocity squared $\Omega_I^2$~\subref{fig:Tc-DBC-Om} and the linear boundary velocity squared $v_I^2$~\subref{fig:Tc-DBC-V}. Results are presented for several lattice sizes $N_t
\times N_z \times N_s^2$ with DBC. Lines correspond to simple quadratic fits $T_c(\Omega_I)/T_c(0)=1-C_2\Omega_I^2$ and $T_c(v_I)/T_c(0)=1-B_2v_I^2/c^2$}
\label{fig:Tc-DBC-var}
\end{figure}

In Fig.~\ref{fig:D-Om} we present the Polyakov loop and its susceptibility as functions of the ratio $T/T_c(0)$ for the lattice size $8\times24\times49^2$. One important difference between DBC and other two boundary conditions is that the Polyakov loop does not go to zero in the confinement phase: it is a consequence of the explicit symmetry breaking. Nevertheless, there is a clear inflection point for the Polyakov loop, as well as the peak for the susceptibility. Using the standard Gaussian fit (\ref{eq:chi_fit}) we determine the critical temperature from the Polyakov loop susceptibility peak, which is shown in Fig.~\ref{fig:Tc-DBC-Om}. In Fig.~\ref{fig:Tc-DBC-V} we present, similarly to other BC, the ratio $T_c(v_I)/T_c(0)$ as a function of the (imaginary) boundary velocity $v_I$.

\begin{figure}[htb]

\includegraphics[width=.48\textwidth]{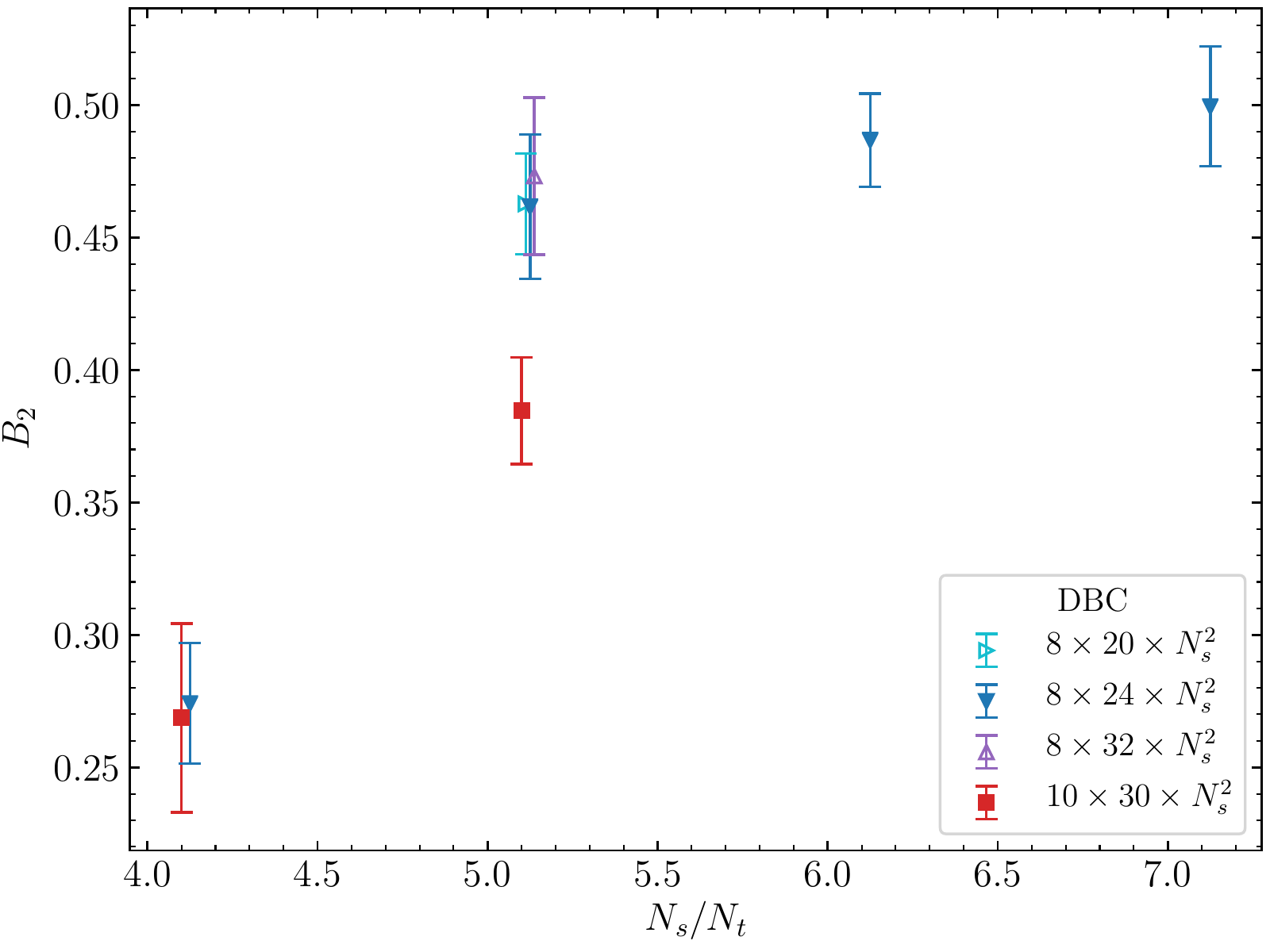}
\caption{The coefficient $B_2$ in Eq.~(\ref{eq:Tc-fitimagv}) versus the ratio $N_s/N_t$ for several lattice sizes with DBC.}\label{fig:B2-DBC}
\end{figure}

\begin{figure*}
\subfigure[]{\label{fig:Lxy_Dir-comp-0}
\includegraphics[width=0.48\textwidth]{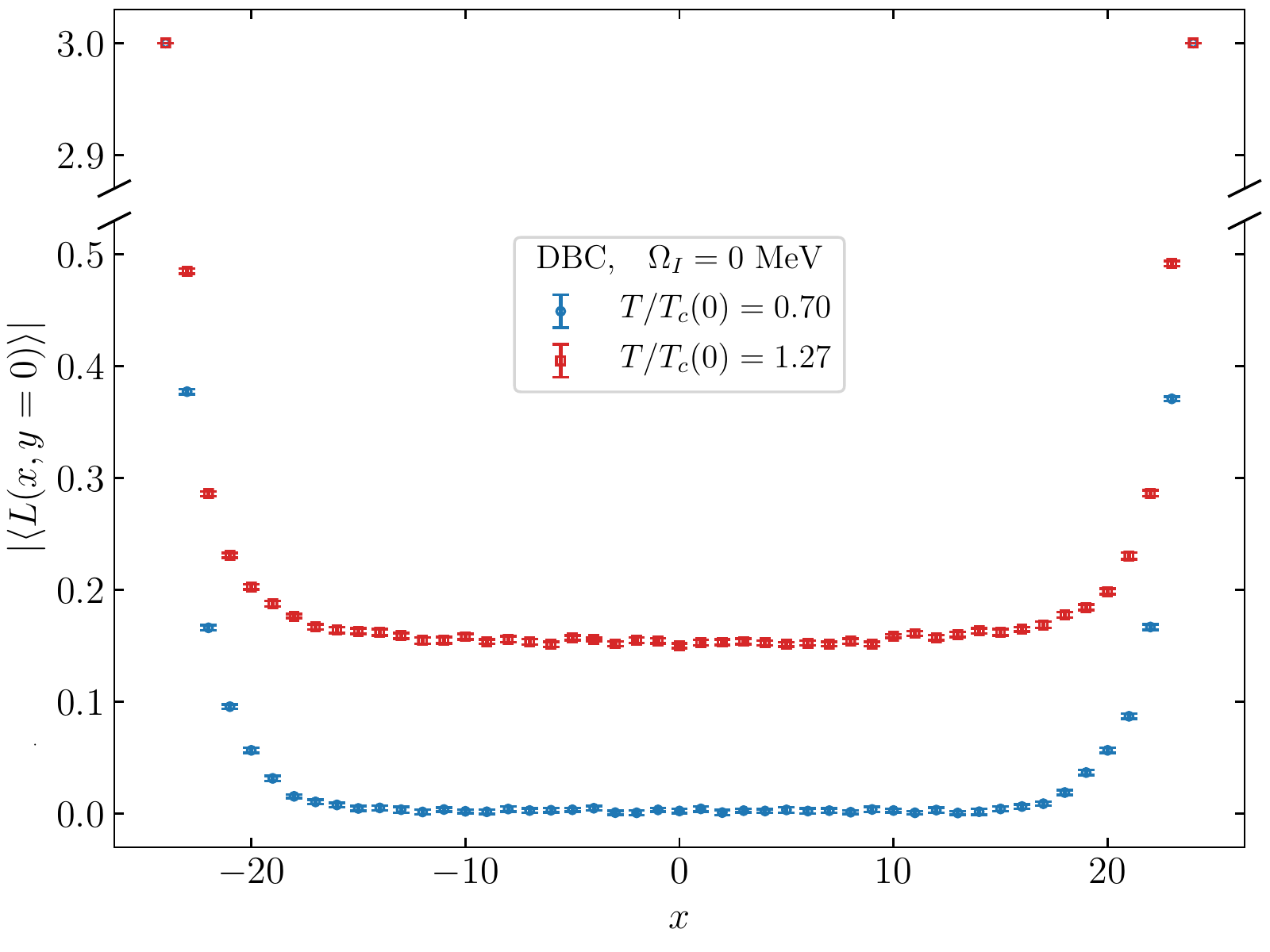}
}
\hfill
\subfigure[]{\label{fig:Lxy_Dir-comp-24}
\includegraphics[width=0.48\textwidth]{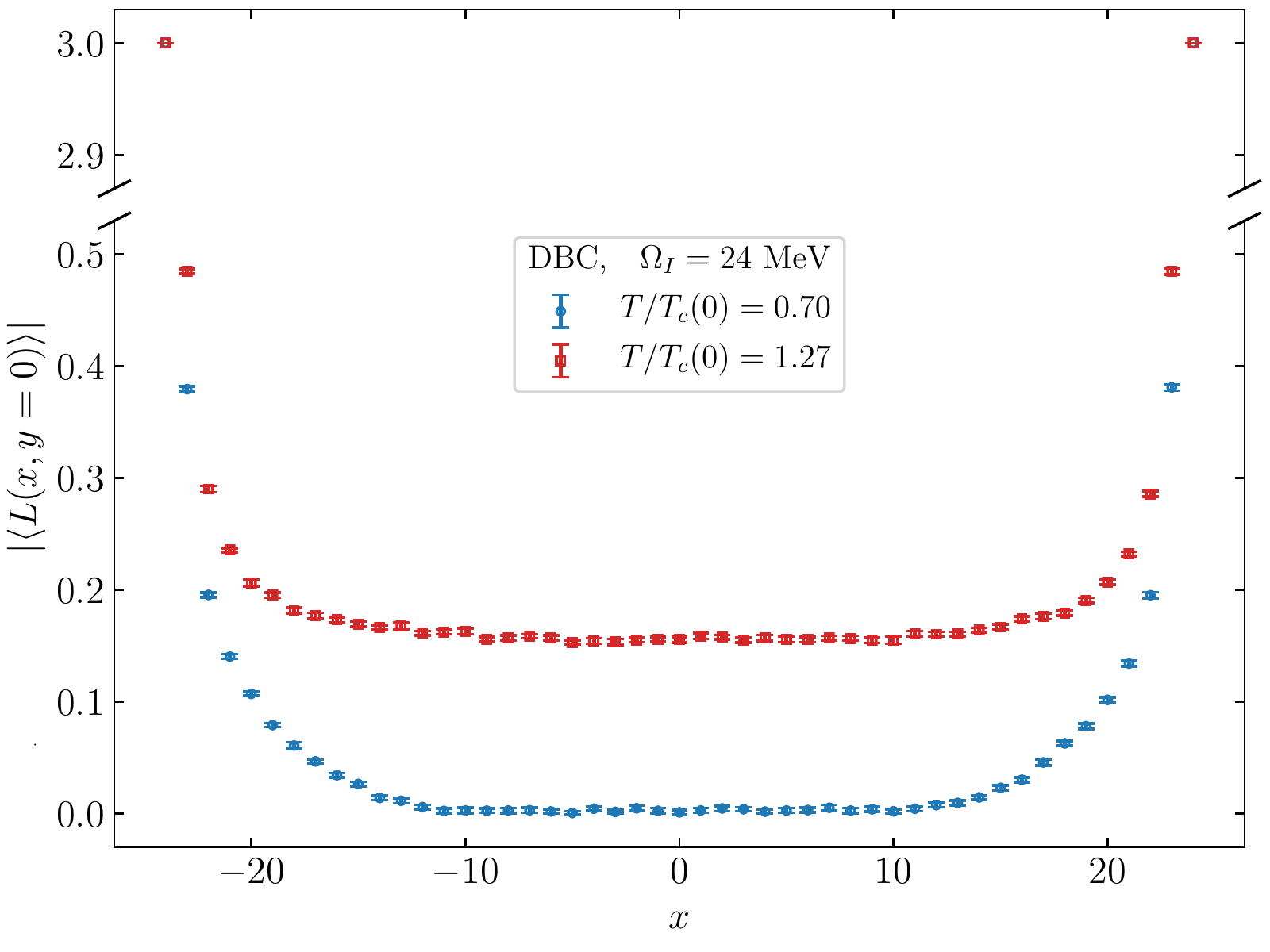}
}
\caption{The Polyakov loop $| \langle L(x,y=0) \rangle |$ as a function of coordinate $x$ for DBC and $\Omega_I = 0$ MeV \subref{fig:Lxy_Dir-comp-0}, $\Omega_I = 24$ MeV \subref{fig:Lxy_Dir-comp-24}. The results were obtained on the lattice $8\times 24 \times 49^2$ for two temperatures: $T/T_c(0)=0.70$ in the confinement phase and $T/T_c(0)=1.27$ in the deconfinement phase.
}
\label{fig:Lxy_Dir-comp}
\end{figure*}

The behaviour of the critical temperature $T_c$ in rotating gluodynamics with DBC is completely analogous to OBC and PBC:
\begin{itemize}
    \item The ratio $T_c(\Omega_I)/T_c(0)$ is described with good accuracy by a function $T_c(\Omega_I)/T_c(0)=1-C_2\Omega_I^2$, with $C_2>0$. After analytical continuation we draw a conclusion that {\it with DBC the critical temperature of the confinement/deconfinement transition increases with angular velocity}.

    \item The coefficient $C_2$ has very weak dependence on the lattice spacing and lattice size $N_z$ but significantly changes with the size $N_s$.
    If one takes instead $\Omega_I$ the linear boundary velocity $v_I$: $T_c(v_I)/T_c(0)=1-B_2v_I^2$, the ratio $T_c(v_I)/T_c(0)$ exhibits significantly smaller dependence on the lattice size $N_s$ (see Fig.~\ref{fig:Tc-DBC-V}). Looking at Fig.~\ref{fig:Tc-DBC-V} one can see that due to large uncertainties the divergence of lines with different $N_s$ is quite significant. The coefficient $B_2$ is presented in Fig.~\ref{fig:B2-DBC}. From this figure it is seen that similarly to OBC, for sufficiently large $N_s$ the contribution of the boundary is suppressed and the $B_2$ goes to plateau with the value $B_2 \sim 0.5$. 
 
    \item In Fig.~\ref{fig:Lxy_Dir-comp} we show the dependence of the Polyakov loop on the spatial coordinate for the angular velocities $\Omega_I=0~\mbox{MeV}, 24$~MeV. The results are shown for two temperatures: $T/T_c(0)=0.70$ in the confinement phase and $T/T_c(0)=1.27$ in the deconfinement phase. 
    
    DBC fix the value of the Polyakov loop $L(x,y)=3$ on the boundary. Similarly to OBC, the effect of boundary conditions is screened and at sufficiently large distances from the boundary the Polyakov loop tends to a constant, which is zero in the confinement phase and nonzero for the deconfinement.
    Comparing Fig.~\ref{fig:Lxy_Dir-comp-0} and Fig.~\ref{fig:Lxy_Dir-comp-24} one can notice that rotation induces additional inhomogeneity of Polyakov loop but its influence is much weaker as compared to that due to the BC.     
\end{itemize}

\section{Discussion and conclusion}

In this paper we addressed the question how rotation influences the confinement/deconfinement transition in gluodynamics within lattice simulation. 
We perform the simulation in the reference frame which rotates with the system under investigation. In this reference frame rotation is reduced to external gravitational field. Having constructed the action of lattice gluodynamics in external gravitational field we found that this action is spoiled by sign problem and direct application of Monte Carlo importance sampling is not possible. To overcome this problem we conducted our study for sufficiently small imaginary angular velocities~$\Omega_I$ and the results were analytically continued to real values of angular velocity~$\Omega$. Our results suggest that this approach is applicable for the values of angular velocity which are characteristic for heavy-ion collision experiments. 

Because of the causality, the simulation of rotating gluodynamics has to be carried out with boundary conditions (BC).  It is important to stress that BC are important part of all approaches aimed at studying of rotating quark-gluon plasma rather than a lattice artifact. The results obtained within any approach depend on BC. In order to study this dependence in our paper we applied various BC. In particular, our simulations were carried out with open boundary conditions (OBC), periodic boundary conditions (PBC), Dirichlet boundary conditions (DBC). In our paper we are mainly focused on the influence of rotation on the critical temperature of the confinement/deconfinement transition. The critical temperature was determined from the peak of Polyakov loop susceptibility. 

The results obtained in our work allow us to state that after the analytical continuation the $T_c(\Omega)$ can be well described  by a simple quadratic function
\begin{equation}\label{eq:Tc-fitimag_concl}
    \frac{T_c(\Omega)}{T_c(0)} = 1 + C_2 \Omega^2,
\end{equation}
with $C_2>0$ for all BC and all lattice parameters used in the simulations. From this result we draw the main conclusion of our paper {\it the critical temperature of the confinement/deconfinement phase transition grows with increasing angular velocity.} This conclusion does not depend on BC and we believe that this is universal property of gluodynamics. 

The magnitude of the coefficient $C_2$ depends on BC. For each boundary condition used in the simulations, the $C_2$ does not depend on the lattice size along the rotation axis $N_z$, has weak dependence on the lattice spacing, but it has strong dependence on the lattice size perpendicular to the rotation axis $N_s=N_x=N_y$. The leading dependence of formula~\eqref{eq:Tc-fitimag_concl} on the $N_s$ can be captured if one rewrites it in terms of the linear velocity $v$ on the boundary $v = \Omega\, (N_s-1)\, a/2$ as follows
\begin{equation}\label{eq:Tc-fitrealv_concl}
    \frac{T_c(v)}{T_c(0)} = 1 + B_2 \frac{v^2}{c^2}.
\end{equation}
In last formula the coefficient $B_2$ weakly depends on $N_s$. We believe the possibility to describe our results for all $N_s$ by universal formula (\ref{eq:Tc-fitrealv_concl}) rather than formula (\ref{eq:Tc-fitimag_concl}) originates from the following fact. For the thermodynamics of rotating system the most important physical object is the field of velocities and how close this field approaches the speed of light. This property is determined by the product $\Omega a (N_s-1)/2$ but not the $\Omega$ alone. 
In addition we also found that for OBC $B_2 \sim 0.7$, for PBC $B_2 \sim 1.3$ and for DBC $B_2 \sim 0.5$. So, although the values of the $B_2$ are close to each other, we still see the dependence of our quantitative results on BC. 

One might suspect that raise of the critical temperature with 
rotation is related to space-dependent temperature due to Ehrenfest–Tolman effect. 
However, we believe that this is not the case. On the contrary Ehrenfest–Tolman effect would 
lead to decrease of the critical temperature with rotation. This can be 
seen as follows. When one moves from the rotation axis to 
to the boundary, the space-dependent temperature increases as compared to the rotation axis.
As the result one needs lower temperature at the rotation axis to realize the confinement/deconfinement transition. Notice that this conclusion is in agreement with 
the result of papers \cite{Chernodub:2020qah}, where the dependence of the critical temperature 
on the angular velocity was linked to the Ehrenfest–Tolman effect. However, our results demonstrate the opposite behaviour: we observe the increase 
of the critical temperature with rotation.

In paper~\cite{Chernodub:2020qah} it was proposed that rotation might lead to mixed inhomogeneous phase when the matter is in the confinement phase close to the center of rotation whereas the deconfinement takes place close to the boundary. Unfortunately for lattice parameters used in our study we have not found such state. Probably a more thorough study is required to answer the question whether such a phase is realized in rotating gluodynamics. 

The authors of papers~\cite{Chen:2020ath, Chernodub:2020qah, Fujimoto:2021xix} studied the confinement/deconfinement transition in rotating QCD within phenomenological models. The results of these works indicate that the critical temperature decreases with the angular velocity, which is in disagreement with the results of our work. The origin of this disagreement is not yet clear, in the future we plan further investigation of this problem.

\begin{acknowledgments}
We would like to thank Valentin Zakharov, Oleg Teryaev, Maxim Chernodub, Vitaly Bornyakov for useful discussions.
This work was supported by RFBR grants 18-02-40126.
This work has been carried out using computing resources of the Federal collective usage center Complex for Simulation and Data Processing for Mega-science Facilities at NRC ``Kurchatov Institute'', \url{http://ckp.nrcki.ru/}; 
the cluster of Institute for Theoretical and Experimental Physics and the Supercomputer  ``Govorun'' of Joint Institute for Nuclear Research.
\end{acknowledgments}

\medskip

\appendix
\section{Finite volume effects on the lattices with Dirichlet and open boundary conditions}\label{sec:app-Tc0}  

In this section we are going to address the question how boundary conditions (BC) considered in this paper influence the critical temperature for gluodynamics without rotation. 
For all BC used in our paper we apply periodic boundary conditions for gluon fields in $z$- and $t$-directions. What concerns the $x$- and $y$-directions we used periodic (PBC), Dirichlet (DBC) and open (OBC) boundary conditions (see Section~\ref{sec:II-lattice} for details). Since PBC are common for lattice simulations, there are a lot of papers where the volume dependence of critical temperature within PBC was studied (see, for instance, \cite{Boyd:1996bx}). For this reason, in Appendix we are going to focus on DBC and OBC only. 

\subsection{Open boundary conditions}

In this section we consider OBC. The critical temperature is determined from the peak of the Polyakov loop susceptibility
(\ref{eq:polyakov_chi}). The results of this calculation for different lattices can be found in Table~\ref{tab:Tc-0}. It is seen that for all lattices presented in this table the critical temperature for OBC is larger than that in gluodynamics with PBC $T_c/\sqrt \sigma = 0.6383(55)$~\cite{Boyd:1996bx}.\footnote{Notice that in this and next section we compare our results for the critical temperatures with the result obtained in the infinite volume limit on the lattices with PBC and $N_t=8$~\cite{Boyd:1996bx}. This is because in Appendix we do not take continuum limit and conduct the simulations on $N_t=8$ lattices.} This feature of OBC can be understood as follows. In OBC the plaquettes outside the lattice volume
are excluded. This can be considered as if one puts these plaquettes to unity what leads to zero lattice action.
 So, physically this can be interpreted as if the studied volume is attached to classical (without quantum fluctuations) zero temperature Yang-Mills theory. From this perspective the regions near the boundary have lower temperature than the regions remote from the boundary. So, in order to observe confinement/deconfinement transition with OBC one has to heat the system to larger temperature as compared to homogeneous gluodynamics. 

To study the infinite volume limit we fit our data for the lattices $8 \times N_s^3$ by the function
\begin{equation}\label{eq:Tc-Ns}
T_c(N_s/N_t) = T_0 + A (N_t/N_s)^3 
\end{equation}
where $T_0$ corresponds to the infinite volume limit. The fit gives $T_0/\sqrt{\sigma}=0.6420(18)$.
This value is in reasonable agreement with that obtained in the infinite volume limit for the PBC: $T_0/\sqrt{\sigma} = 0.6383(55)$~\cite{Boyd:1996bx}.

From these results one can draw a conclusion that finite volume effects for OBC enhance the critical temperature. As one increases the volume, the effect of BC becomes screened and the critical temperature for OBC approaches to its acknowledged value~\cite{ Boyd:1996bx}. The screening of the boundary is well seen in Fig.~\ref{fig:Lxy_open-comp-0}.  

\begin{table}[h]
\caption{The critical temperature for non-rotating lattices with OBC and DBC. These results to be compared with the critical temperature of gluodynamics with PBC: $T_c/\sqrt \sigma = 0.6383(55)$~\cite{Boyd:1996bx}.}\label{tab:Tc-0}
\begin{ruledtabular}
\begin{tabular}{rll}
      & \mc{OBC} & \mc{DBC} \\
    \mc{Lattice} & \mc{$T_c/\sqrt{\sigma}$} & \mc{$T_c/\sqrt{\sigma}$} \\
    \colrule\noalign{\smallskip}
    $8\times 40^3$ &  0.6983(10) &  0.5764(8) \\
    $8\times 48^3$ &  0.6755(6) &  0.5886(9)  \\
    $8\times 56^3$ &  0.6623(10) &  0.5940(11) \\
    $8\times 64^3$ &  0.6537(10) &  0.6021(9) 
\end{tabular}
\end{ruledtabular}
\end{table}

\subsection{Dirichlet boundary conditions}

In this section we consider DBC. The results for the critical temperature calculation with DBC for different lattices can be found in Table~\ref{tab:Tc-0}. It is seen that for all lattices presented in this table the critical temperature for DBC is smaller than that in gluodynamics  with PBC~\cite{Boyd:1996bx}. In this sense OBC and DBC influence to the system in the opposite way. The decrease of the critical temperature in DBC as compared to PBC gluodynamics can be understood as follows. The link variables on the boundary in DBC equal unity, i.e. after taking the trace over colors the Polyakov loop on the boundary equals 3. So, the $\mathbb{Z}_3$ symmetry of lattice gluodynamics is explicitly broken on the boundary. In gluodynamics breaking of $\mathbb{Z}_3$ symmetry is the property of high temperature phase. For this reason in DBC the boundary can be considered as a seed of high temperature phase. On these grounds, one can expect that the confinement/deconfinement transition takes place at smaller critical temperature. 

To study the infinite volume limit we fit our data for the lattices $8 \times N_s^3$ by 
the same function~\eqref{eq:Tc-Ns},
where $T_0$ corresponds to the infinite volume limit. The fit gives $T_0/\sqrt{\sigma}=0.6086(22)$.
This value is in reasonable agreement with that obtained in the infinite volume limit for PBC: $T_0/\sqrt{\sigma} = 0.6383(55)$~\cite{Boyd:1996bx}.

So, finite volume effects in DBC decrease the critical temperature which approaches to that of PBC gluodynamics~\cite{Boyd:1996bx} for sufficiently large volume. Similarly to OBC this behaviour can be explained by screening of the boundary which is well seen on Fig.~\ref{fig:Lxy_Dir-comp-0}. From this figure one can note that the Polyakov loop increases as one approaches to the boundary. This can be interpreted as increase of effective temperature of the regions close to the boundary. 

To summarize of both sections of Appendix, our results allows us to state that for sufficiently large volume bulk properties of the system cannot be considerably modified by either OBC or DBC. We believe that this conclusion remains to be true even for rotating gluodynamics if the angular velocity is not too large.

\section{Lattice parameters used in the simulations}\label{sec:app-list}

In our paper we  performed  numerical  simulations for the  lattice sizes and values of imaginary angular velocity listed in Tab.~\ref{tab:listOmega}.
For each lattice size and angular velocity we changed the temperature through the variation of the $\beta$. To set the physical scale we used the relation between lattice spacing and inverse lattice coupling $(a\sqrt{\sigma})(\beta)$ from Ref.~\cite{Edwards:1997xf} (with the value of  string tension $\sqrt{\sigma} =  440$~MeV).
Simulations are performed with the use of Monte Carlo algorithm, each sweep consists of one heatbath update and two steps of the overrelaxation updates. For each set of parameters, the typical statistics are about 6000--12000 configurations, separated by 20 sweeps. The statistical errors were estimated using jackknife method.

\begin{table}[h]
\caption{The list of lattice parameters used in the simulations with OBC, PBC and DBC .}\label{tab:listOmega}
\begin{ruledtabular}
\begin{tabular}{rlrlrl}
    \multicolumn{2}{c}{OBC} & \multicolumn{2}{c}{PBC} & \multicolumn{2}{c}{DBC}\\
    \cline{1-2} \cline{3-4} \cline{5-6}\noalign{\smallskip}
    \mc{Lattice} & \mc{$\Omega_I$, MeV} & \mc{Lattice} & \mc{$\Omega_I$, MeV} & \mc{Lattice} & \mc{$\Omega_I$, MeV}\\
    \colrule\noalign{\smallskip}
    $8\times 20\times 25^2$ & 0, 30, 45, 60, 75, 90 & $8\times 24\times 25^2$ & 0, 15, 30, 45, 50 & $8\times 24\times 33^2$ & 0, 15, 30, 45 \\
    $8\times 24\times 25^2$ & 0, 30, 45, 60, 75, 90 & $8\times 24\times 33^2$ & 0, 10, 20, 30, 38 &  $8\times 20\times 41^2$ & 0, 10, 20, 30, 36 \\
    $8\times 30\times 25^2$ & 0, 30, 45, 60, 75, 90 &  $8\times 24\times 41^2$ & 0, 12, 20, 24, 30 & $8\times 24\times 41^2$ & 0, 12, 24, 30, 36 \\
    $8\times 24\times 29^2$ & 0, 15, 30, 45, 60, 75 & $8\times 24\times 49^2$ & 0, 10, 15, 20, 24 & $8\times 32\times 41^2$ & 0, 12, 24, 30, 36  \\
    $8\times 24\times 33^2$ & 0, 15, 30, 45, 50 & $10\times 24\times 31^2$ & 0, 15, 30, 45, 50 & $8\times 24\times 49^2$ & 0, 10, 20, 24, 30 \\
    $8\times 24\times 41^2$ & 0, 12, 24, 30, 36 & $10\times 30\times 31^2$ & 0, 15, 30, 45, 50 & $8\times 24\times 57^2$ & 0, 10, 15, 20, 24 \\
    $8\times 32\times 41^2$ & 0, 12, 24, 30, 36 & $10\times 36\times 31^2$ & 0, 15, 30, 45, 50 & $10\times 30\times 41^2$ & 0, 15, 30, 45 \\
    $8\times 24\times 49^2$ & 0, 10, 20, 24, 30 & $10\times 30\times 41^2$ & 0, 10, 20, 30, 38 & $10\times 30\times 51^2$ & 0, 12, 24, 30, 36  \\
    $8\times 24\times 57^2$ & 0, 10, 15, 20, 24 & $10\times 30\times 51^2$ & 0, 12, 20, 24, 30 &\\
    $10\times 30\times 31^2$ & 0, 30, 45, 60, 75, 90 & $12\times 36\times 37^2$ & 0, 15, 30, 45, 50 & \\
    $10\times 36\times 31^2$ & 0, 30, 45, 60, 90 & $12\times 36\times 49^2$ & 0, 10, 20, 30, 38\\
    $10\times 30\times 41^2$ & 0, 15, 30, 45, 50 \\
    $10\times 30\times 51^2$ & 0, 12, 24, 30, 36 \\
    $12\times 36\times 37^2$ & 0, 30, 45, 60, 75, 90
\end{tabular}
\end{ruledtabular}
\end{table}

\medskip
\newpage

\end{document}